\documentclass[manuscript]{aastex}

\newcommand{\kms}{km~s$^{-1}$}

\shorttitle{SN~2009bb}

\begin{document}


\title{SN~2009bb: a Peculiar Broad-Lined Type Ic Supernova\altaffilmark{1,2}}


\author{Giuliano~Pignata,\altaffilmark{3,4} 
Maximilian~Stritzinger,\altaffilmark{5,6,7}
Alicia~Soderberg,\altaffilmark{8}
Paolo~Mazzali,\altaffilmark{9,10,11}
M.~M.~Phillips,\altaffilmark{5}
Nidia Morrell,\altaffilmark{5}
J.~P.~Anderson,\altaffilmark{4}
Luis~Boldt,\altaffilmark{5}
Abdo~Campillay,\altaffilmark{5}
Carlos~Contreras,\altaffilmark{5,12}
Gast\'on~Folatelli,\altaffilmark{4}
Francisco~F\"orster,\altaffilmark{4}
Sergio~Gonz\'alez,\altaffilmark{5}
Mario~Hamuy,\altaffilmark{4}
Wojtek~Krzeminski,\altaffilmark{5}
Jos\'e~Maza,\altaffilmark{4}
Miguel~Roth,\altaffilmark{5}
Francisco~Salgado,\altaffilmark{5}
Emily~M.~Levesque,\altaffilmark{13}
Armin~Rest,\altaffilmark{14}
J.~Adam~Crain,\altaffilmark{15}
Andrew~C.~Foster,\altaffilmark{15}
Joshua~B.~Haislip,\altaffilmark{15}
Kevin~M.~Ivarsen,\altaffilmark{15}
Aaron~P.~LaCluyze,\altaffilmark{15}
Melissa~C.~Nysewander,\altaffilmark{15}
Daniel~E.~Reichart\altaffilmark{15}
}


\altaffiltext{1}{This paper includes data gathered with the 
6.5-m Magellan Telescopes located at Las Campanas Observatory, Chile.}
\altaffiltext{2}{Based on observations obtained at the Gemini Observatory, Cerro Pachon, Chile (Gemini Programs GS-2009A-Q-17 and GS-2009A-Q-43).}
\altaffiltext{3}{Departamento de Ciencias Fisicas, Universidad Andres Bello, Avda.
Republica 252, Santiago, Chile}
\altaffiltext{4}{Departamento de Astronom\'ia, Universidad de Chile, Casilla 36-D, Santiago, Chile.}
\altaffiltext{5}{Las Campanas Observatory, Carnegie Observatories,
  Casilla 601, La Serena, Chile.}
\altaffiltext{6}{Dark Cosmology Centre, Niels Bohr Institute, University 
of Copenhagen, Juliane Maries Vej 30, 2100 Copenhagen \O, Denmark.}
\altaffiltext{7}{The Oskar Klein Centre, Department of Astronomy, Stockholm University, AlbaNova, 10691 Stockholm, Sweden.} 
\altaffiltext{8}{Harvard-Smithsonian Center for Astrophysics,
60 Garden Street, Cambridge, MA 02138, USA.}
\altaffiltext{9}{Max-Planck-Institut f\"ur Astrophysik, Karl-Schwarzschild- Strasse 1, 85741 Garching, Germany.}
\altaffiltext{10}{Scuola Normale Superiore, Piazza Cavalieri 7, 56127 Pisa, Italy.}
\altaffiltext{11}{INAF Oss. Astron. Padova, vicolo dellOsservatorio 5, 35122 Padova, Italy}
\altaffiltext{12}{Centre for Astrophysics \& Supercomputing, Swinburne University of Technology, P.O. Box 218, Victoria 3122, Australia.}
\altaffiltext{13}{Institute for Astronomy, University of Hawaii, 2680 Woodlawn Drive, Honolulu, HI 96822, USA}
\altaffiltext{14}{Department of Physics, Harvard University, 17 Oxford Street, Cambridge, MA 02138, USA}
\altaffiltext{15}{University of North Carolina at Chapel Hill, Campus Box 3255, Chapel Hill, NC 27599-3255, USA.} 

\begin{abstract}
Ultraviolet, optical, and near-infrared  photometry and optical spectroscopy of the broad-lined Type~Ic supernova (SN) 2009bb are presented, following the flux evolution from $-$10 to $+$285 days past $B$-band maximum. Thanks to the very early discovery, it is possible to place tight constraints on the SN  explosion epoch. The expansion velocities measured from near maximum spectra are found to be only slightly smaller than   those measured from spectra of the prototype broad-lined SN~1998bw associated with GRB~980425. 
Fitting an analytical model to the pseudo-bolometric light curve of SN~2009bb suggests that 4.1$\pm$1.9 M$_{\sun}$ of material was ejected with 
0.22 $\pm$0.06 M$_{\sun}$ of it being $^{56}$Ni. The resulting kinetic energy is 1.8$\pm$0.7$\times 10^{52}$ erg. This, together with an absolute peak magnitude of M$_{B}$ $=$ $-$18.36$\pm$0.44, places SN~2009bb on the energetic and luminous end of the broad-lined Type~Ic (SN~Ic) sequence. 
Detection of helium in the early time optical spectra accompanied with
strong radio emission, and high metallicity of its environment makes
SN~2009bb a peculiar object.
Similar to the case for GRBs, we find that the bulk explosion parameters of SN~2009bb cannot account for the copious energy coupled to relativistic ejecta, and conclude that another energy reservoir (a central engine) is required to power the radio emission.
Nevertheless, the analysis of the SN~2009bb nebular spectrum suggests that the  failed GRB detection is not imputable to a large angle between the line-of-sight and the GRB beamed radiation. Therefore, if a GRB was produced during the SN~2009bb explosion, it was below the threshold of the current generation of $\gamma$-ray instruments.

\end{abstract}


\keywords{galaxies: individual (NGC 3278) --- supernovae: general
  --- supernovae: individual (SN~2009bb)}

\section{Introduction}

\begin{figure}
\epsscale{.80}
\plotone{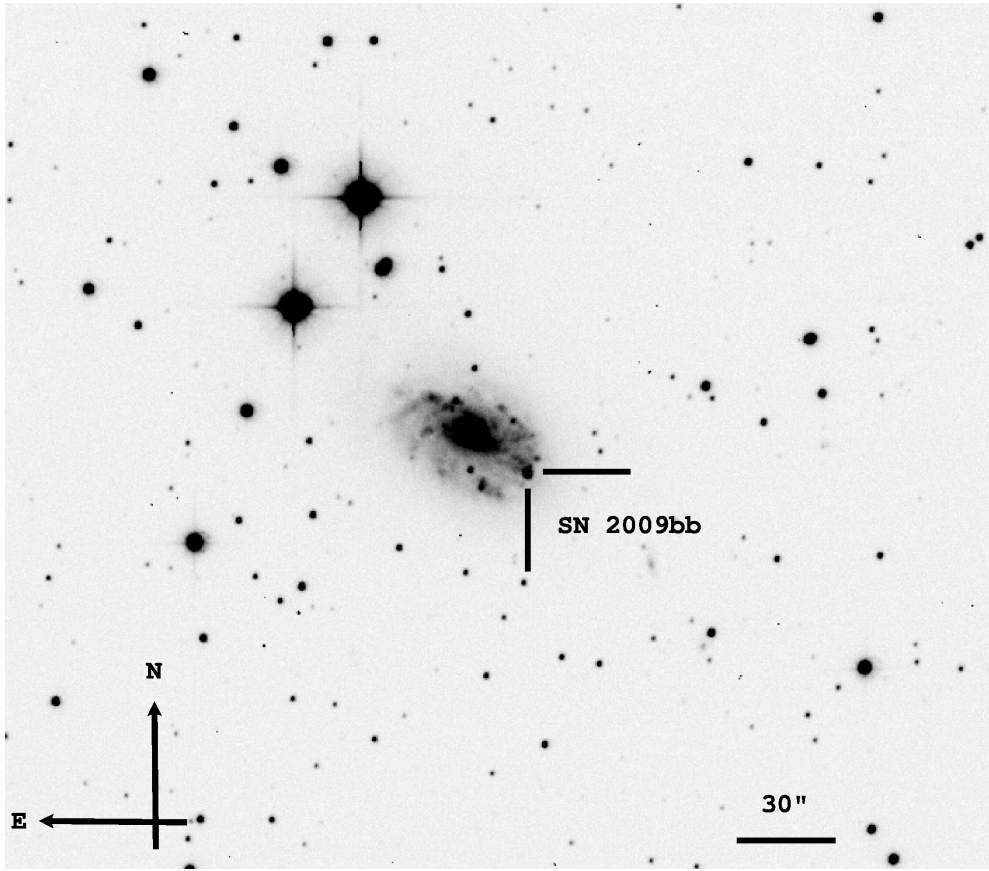}
\caption{Finder chart for SN~2009bb obtained from a $V$ band image acquired on 2009 March 30 with the 1.0~m Swope telescope at Las Campanas Observatory.}
\label{fig1.1}
\end{figure}

A decade has passed since the discovery of the 
association between the long-duration Gamma Ray Burst (GRB)~980425 and 
the broad-lined Type~Ic SN~1998bw \citep{Galama98,Patat01}.
This realization led to a  renewed interest in the study of stripped-envelope core-collapse supernovae (SNe).
In the case of SN~1998bw, the high kinetic energy inferred from the expansion velocity 
coupled with an exceedingly high luminosity  
and strong radio emission set it apart from all other previously-observed core-collapse
 SNe. 
 Since then these attributes have been shown also by other GRB-related SNe like 
 SN~2003lw \citep{Malesani04,Gal-Yam04} and SN~2003dh \citep{Stanek03,Hjorth03,Matheson03}. The X-ray flash SN~2006aj \citep{Pian06,Modjaz06,Sollerman06,Ferrero06,Kocevski07}, was not as extreme as the previous SNe, nevertheless it shows expansion velocity much higher then normal Type Ic SNe. 

In addition to these GRB related, broad-lined SNe~Ic, there have also been a number of other 
discovered broad-lined events that are apparently not associated with a GRB, e.g. 
 SN~1997ef, \citep{Matheson01},  SN~2002ap \citep{Gal-Yam02,Foley03,Yoshii03}, SN~2003jd \citep{Valenti08} and SN~2007ru \citep{Sahu09}. 
Recently two other broad-lined events have been published
that underscore the heterogeneous nature of this family of SNe. In particular, 
early phase spectra of the broad-lined Type~Ic SN~2007bg exhibit evidence of
{\em helium} \citep{Young10}, while \citet{Hamuy09} have presented the first 
case of a  {\em hydrogen-rich} broad-lined Type~IIb SN~2003bg.

In this paper we present ultraviolet, optical and near-infrared photometry and optical spectroscopy of SN~2009bb. This object was discovered by the CHilean Automated Supernova sEarch CHASE \citep{Pignata09a}  on 2009 March 21.11 UT with the Panchromatic Robotic Optical Monitoring and Polarimetry Telescope (PROMPT) 3 at the Cerro Tololo Inter-American Observatory (CTIO).\footnote[16]{Cerro Tololo Inter-American Observatory, Kitt Peak National Observatory,
National Optical Astronomy Observatories, operated by the Association
of Universities for Research in Astronomy, Inc., (AURA), under cooperative
agreement with the National Science Foundation.}
The SN  is located (see Figure~\ref{fig1.1}) at  $\alpha$ $=$ $10^{h}31^{m}33\fs87$  and 
$\delta$ $=$ $-39\degr57\arcmin30.0\arcsec$ (equinox J2000) \citep{Pignata09b},   
which is about 17\farcs0 west and 13\farcs5 south of the center of the host galaxy NGC 3278.
SN~2009bb was not visible in an unfiltered   CHASE 
image  (mag $<$ 19.2) obtained two days prior to the discovery image on 2009 March 19.20  UT.
We were therefore able to use this image in combination with the first follow-up images
to obtain a tight constraint 
on the explosion epoch (see section 2.3).

As this was an extremely young SN, an intensive
follow-up campaign was organized using the PROMPT telescopes  \citep{Reichart05} and the facilities available to the Carnagie Supernova Project (CSP; Hamuy et al. 2006) at Las Campanas Observatory (LCO). A week after discovery,  
 \citet{Stritzinger09} classified SN~2009bb as a broad-lined SN~Ic.

Radio and X-ray observations were obtained with the VLA and Chandra telescope.
An analysis of these data indicates that SN~2009bb was associated with strong 
radio emission and mild X-ray emission \citep{Soderberg10}. 
These findings are suggestive of the presence of a  relativistic outflow 
typically associated with GRB-related SNe. 

The explosion site of SN~2009bb has been  studied in detail by  \citet{Levesque10a} who  showed that contrary to other jet-driven SN explosions, which to date have always been identified with  metal poor environments [see \citet{Modjaz08} for a review], the explosion site of SN~2009bb exhibits a 
 metallicity between  1.7 to 3.5 Z$_{\sun}$. \citet{Levesque10b} also found a super-solar metallicity for the environment of the long-duration GRB~020819.
 These findings present a challenge to the theoretical framework developed to explain GRBs \citep{Woosley06}. However, \citet{Dessart08} have recently proposed a  model in which the progenitor star does not  need to be in a low
 metallicity environment. In this scenario SN~2009bb represents a peculiar object that could open new theoretical developments.

The organization of this article is as follows.
 The photometric data are analyzed in Section~2, and in Section~3 the pseudo-bolometric light curve is used to estimate some physical parameters of SN~2009bb. An analysis of the spectroscopic evolution of SN~2009bb is carried out in Section~4, and Section 5 presents a discussion and summary of the major results. Observation and data reduction techniques are described in Appendix  A.1 and A.2.

\section{Optical and Near-Infrared Photometry}

\begin{figure}
\epsscale{.80}
\plotone{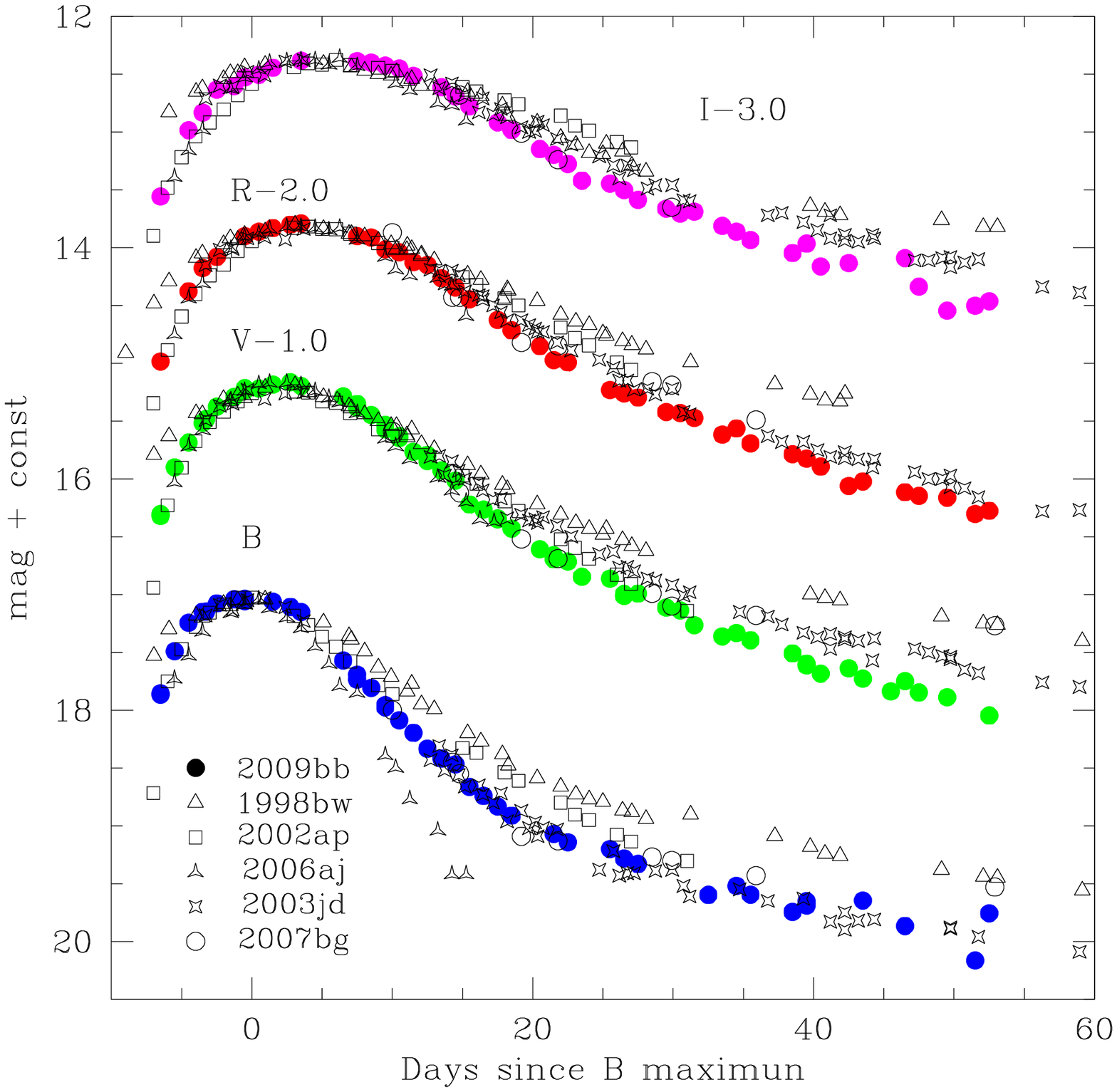}
\caption{$BVRI$ light curves of SN~2009bb. Our data are shown with
filled circles using different colors for different filters.  For
comparison, the light curves of other broad-lined SNe~Ic  are also plotted: SN~1998bw
\citep{Galama98,Patat01}, SN~2002ap \citep{Foley03},  SN~2006aj \citep{Sollerman06,Ferrero06}, SN~2003jd \citep{Valenti08} and SN~2007bg \citep{Young10}. The light curves of the different SNe were shifted to match around maximum. Different colors were shifted by different amount.} 
\label{fig2.1}
\end{figure}

\subsection{Light Curve}

The $BVRI$ photometry of  SN~2009bb is reported in Table~1, while the light curves are shown in Figure~\ref{fig2.1}. Also plotted in the latter figure are photometry of the broad-lined Type~Ic  SNe~1998bw, 2002ap, 2003jd, 2006aj and SN~2007bg. 
From this comparison of light curves, it is clear that in the $VRI$ bands
during the pre-maximum phase, SN~2009bb, SN~2006aj and SN~2003jd show a similar evolution, while SN~1998bw and SN~2002ap  have a slower and faster rise-time, respectively. 
In the $B$ band, SN~2009bb shows, with the exception of  SN~1998bw,  a slower rise than all the other SNe included in the plot. 
At post-maximum phases (epoch $>$ $+$20 days), the 
$VRI$ light curves of SN~2009bb decrease in magnitude faster than the other objects, except for SN~2006aj, which appears to be quite similar to SN~2009bb. This is not the case in the $B$ band where the SN~2006aj decline rate is clearly faster than that of SN~2009bb. In the latter band, the object that most resembles SN~2009bb is SN~2003jd, while SN~2002ap and SN~1998bw display much slower decline rates.

\begin{figure}
\epsscale{.80}
\plotone{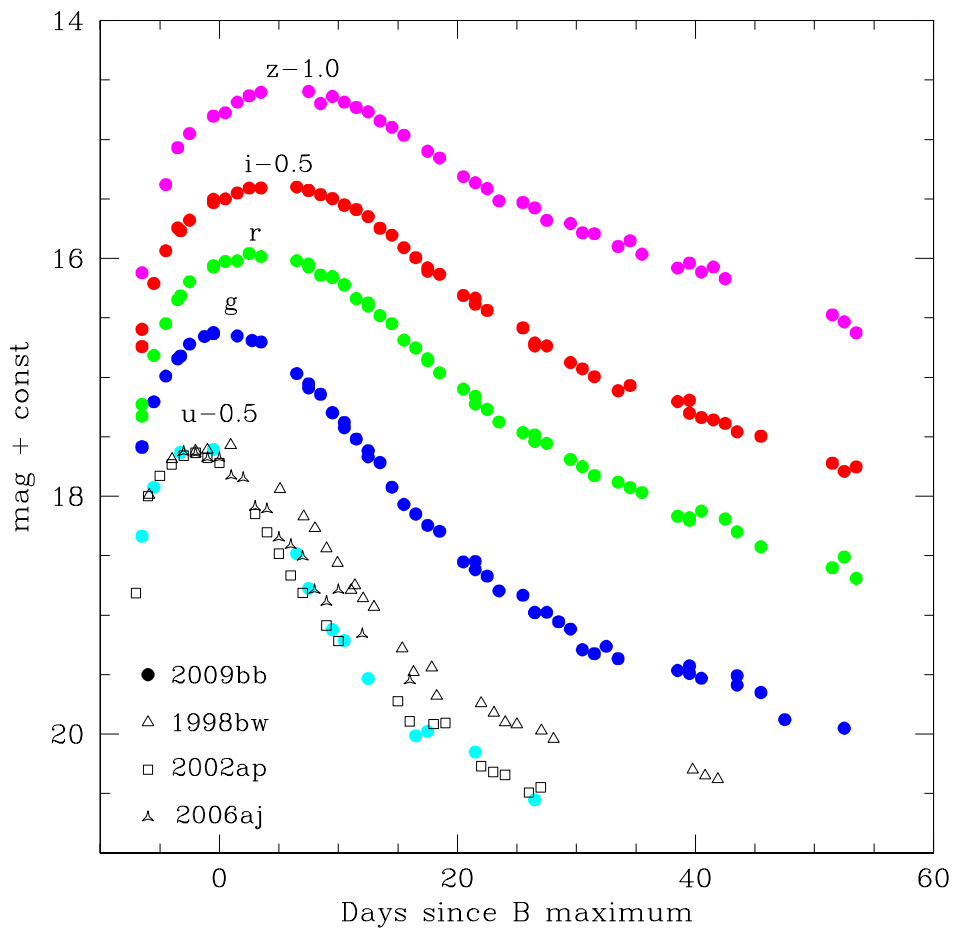}
\caption{$u'g'r'i'z'$ light curves of SN~2009bb. Our data are shown with
filled circles using different colors for different filters.  For
comparison, the $U$ band light curves of SN~1998bw, SN~2002ap and  SN~2006aj are included. The light curves of the different SNe were shifted to match around maximum. The bibliographic sources for the light curves are the same as those given in Figure~\ref{fig2.1}.}
\label{fig2.2}
\end{figure}

The $u'g'r'i'z'$ photometry of  SN~2009bb is reported in Table~2, while the light curves are shown in Figure~\ref{fig2.2}. 
These are the first-ever observations of a broad-lined SN~Ic in the Sloan bands. 
For comparison in the plot, we also include $U$-band light curves of SN~1998bw, SN~2002ap and SN~2006aj. 
The $u'$ light curve of SN~2009bb most resembles the $U$-band observations of SN~2002ap, but again declines more rapidly than SN~1998bw. It should be stressed that Figure~\ref{fig2.2} compares light curves on different photometric systems, and therefore the comparison should be taken with caution.

\begin{figure}
\epsscale{.80}
\plotone{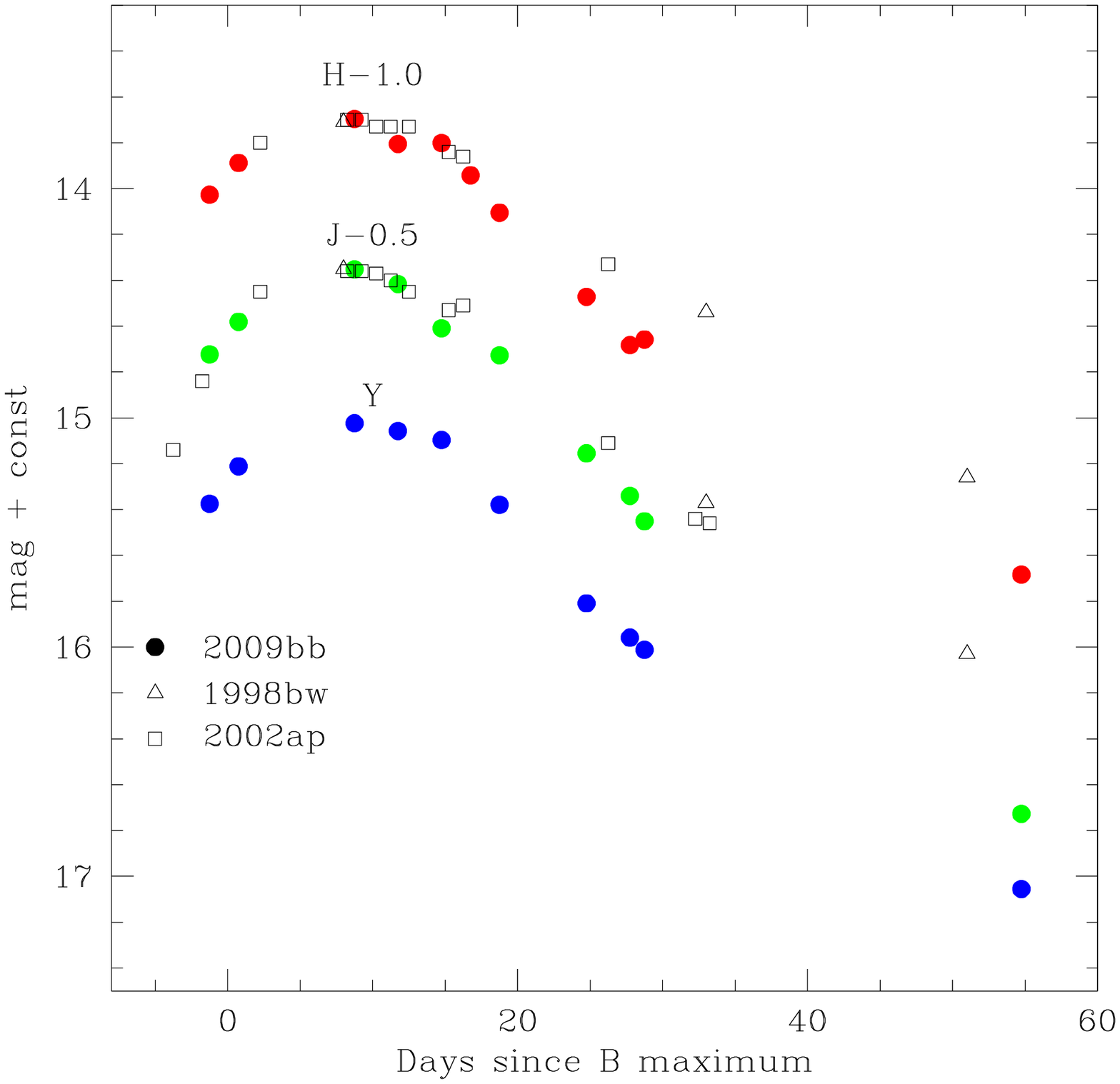}
\caption{$YJH$ light curves of SN~2009bb. Our data are shown with
filled circles using different colors for different filters.  For
comparison, the light curves of SN~1998bw \citep{Patat01} and  SN~2002ap \citep{Yoshii03} are included. }
\label{fig2.3}
\end{figure}

The $YJH$ photometry of  SN~2009bb is reported in Table~3.
In Figure~\ref{fig2.3}, the near-infrared light curves of our SN are displayed together with those of 
SN~1998bw and SN~2002ap. 
As seen in the optical, the post-maximum evolution of SN~2009bb is slightly faster than
that of SN~2002ap. In the case of  SN~1998bw, the difference is larger, with the latter object being $\pm$0.7 and $\pm$0.2 magnitudes brighter than  SN~2009bb in the $J$ and $H$ bands, respectively, at $\sim$50~days past $B$ maximum. Is is worth mentioning that part of this difference could be due to host galaxy contamination affecting the SN~1998bw infrared photometry \citep{Patat01}.
  
For each optical and near-infrared band, the time and value of peak magnitude was estimated from low-order polynomial  fits.
The results are reported in Table~4.  Similar to other SNe~Ib/c, maximum light occurs in the blue 
followed by the red bands. 

In Figure~\ref{fig2.0}, the count rates in the $uvw1$ and $u$ UVOT bands at the position of SN~2009bb are plotted. The fluxes were measured in a fixed aperture and, therefore, are the sum of the light coming from the  SN and the background galaxy. From the plot, it is clear that in the $uvw1$ filter, the background flux is already dominant at $\sim$ 6 days past $B$ maximum light. Assuming that the count rate measured in the last three $uvw1$ observations is entirely due to the background, the SN~2009bb flux in the first observation can be estimated by subtracting the weighted average of the flux in these last   three observations from the total count rate. Adding a zero point for the $uvw1$ band of 17.49 $\pm$ 0.03 \citep{Poole08}, a value of 18.1 $\pm$ 0.1\footnote[17]{Additional source of uncertainties related with the red wing of the $uvw1$ filter are difficult to quantify due to the lack of spectra covering the band. Therefore, they were not included in the error budget} is obtained for the $uvw1$ magnitude of SN~2009bb at $\sim$ 6 days before $B$ maximum light.

\subsection{Color Curves} 

The de-reddened color curves of SN~2009bb are compared in
Figure~\ref{fig2.4} with those of SN~1998bw, SN~2002ap, SN~2006aj,
and  SN~2003jd. In the case of SN~2009bb, a color excess of
E($B-V$) = 0.58$\pm0.07$ is assumed (see Section 3), while for 
SN~1998bw, SN~2002ap, SN~2003jd, and SN~2006aj, values of 
0.06 \citep{Patat01}, 0.08 \citep{Foley03}, 0.14 \citep{Valenti08}, and 0.18 \citep{Sollerman06,Campana06}, respectively, were adopted. It is worth mentioning that for all the latter SNe most of the reddening is produced in the Milky Way. All the authors cited above assumed little or no reddening in the host galaxy due to the absence of strong interstellar Na~I~D absorption in the SNe spectra. Since the correlation between  Na~I~D $EW$ and reddening has been shown to have a significant dispersion \citep{Folatelli10,Olivares10,Stritzinger10}, the  reddening estimates for these SNe should be taken with caution.
To convert the color excesses to corrections for each band, we adopted 
an $R_V$ value of 3.1 \citep{Cardelli89}.

As is seen in Figure~\ref{fig2.4}, the color evolution of broad-lined SNe~Ic show significant
variations. This highlights the inadequacy of color indices to estimate host galaxy extinction for these events.
 The early phase $(B-V)_0$ color evolution of SN~2009bb is seen to be most comparable to that of SN~2003jd,
 while SN~1998bw is   $\sim$0.2 mag redder at early and late epochs, but is similar to SN~2009bb between +10 and +20 days.
In the case of $(V-R)_0$, all SNe have very similar evolution except for SN~2006aj which is $\sim$0.2 redder. 
The $(R-I)_0$ color evolution of SN~2009bb is intermediate between SN~2002ap (bluer) and SN~2003jd (redder).

In Figure~\ref{fig2.5}, we compare the reddening-corrected $(V-J)_0$ and $(V-H)_0$ color curves of SN~2009bb with those of SN~2002ap. The  evolution of the two SNe is similar, but SN~2009bb is $\sim$0.3 and $\sim$0.5 mag  bluer in $(V-J)_0$ and $(V-H)_0$, respectively. 

\subsection{Explosion Date}

The short time interval between the CHASE images with negative and positive detection of SN~2009bb allows us to place strong constraints on its explosion date. 
This is done through the use of a simple equation describing an expanding fireball: 

\begin{equation}
L(t) = K \times (t-t_0)^n,
\label{equa6}
\end{equation}
\noindent where $L$ is the luminosity at time, $t$, $K$ is a constant that 
defines the rate of the rise, n=1.8 \citep{Conley06}, and $t_{0}$ is the time of explosion. 
Applying the equation (\ref{equa6}) to the unfiltered light curve, we estimate 
that the explosion occurred on  JD=2454909.6 $\pm$ 0.6, which corresponds to the time of of our non-detection of SN~2009bb. 

As demonstrated in Figure~\ref{fig2.6}, the curve fits reasonably well the unfiltered light curve (reduced $\chi^2$=3.1), except for the epoch of first detection.

\section{Absolute Luminosity}
To compute the absolute luminosity of SN~2009bb requires  estimates of the host galaxy dust extinction and distance.
One method to estimate the host reddening of SN~2009bb  is to measure the Balmer decrement from
the spectrum of the \ion{H}{2} region located beneath the SN.
To this end we use the two SN~2009bb nebular spectra acquired on  Jan 9th and Feb 3th 2010\footnote[18]{In this spectrum there is not significant SN signal}. In these spectra the signal-to-noise ratio of the \ion{H}{2} region is higher than in the spectra taken at earlier epochs, while the SN flux is lower, allowing for a  more precise  measurement of emission from the Balmer lines.
Obtaining an accurate estimate of the H$_{\alpha}$ and H$_{\beta}$ fluxes requires the removal of the light of the SN and any residual stellar contribution. 
This was accomplished by fitting a cubic spline to the underlying continuum in the spectrum. The resulting pseudo-continuum was then subtracted   from the original spectrum.
 Comparing the measured $H_{\alpha}/H_{\beta}$ flux ratio obtained from  the ``cleaned''  spectra  
 to the values listed in \citet{Osterbrock89} (for case B recombination, T=10000$^{\circ}$ K) and assuming the standard total-to-selective extinction ratio R$_v$ = 3.1, we obtain a total reddening (host+Milky Way) of $E(B-V)_{tot}$ = 0.59$\pm$0.10 and $E(B-V)_{tot}$ = 0.57$\pm$0.10 for the 2010~Jan~9 and 2010~Feb~3 spectra, respectively.
There is another (brighter) \ion{H}{2} region located $\sim$~0.3$^{''}$ North,  $\sim$~1.4$^{''}$ East of
the SN.  A measurement of the  Balmer decrement from its spectrum gives a color excess of $E(B-V)_{tot}$ = 0.56$\pm$0.10, which is in good agreement with that obtained at the SN position.
Given the similarity of the three estimates, we will assume the total reddening suffered 
by SN~2009bb to be the average of the values obtained from its two nebular spectra, i.e. $E(B-V)_{tot}$=0.58$\pm$0.07.

An alternative method often used to estimate the reddening suffered along the line-of-sight to a SN 
is the equivalent width ($EW$) of the interstellar Na~I~D absorption.
The weighted average of the $EW$ measured from seven of our low resolution spectra is  
$EW = 2.16\pm0.12$~\AA.  Taking into account relation between the Na~I~D EW and color excess reported in  \citet{Turatto03},
a value of  $E(B-V)_{\rm host}=0.34$ and  $E(B-V)_{\rm host}=1.06$ is implied for the lower and higher slope, respectively.  This confirms that SN~2009bb was significantly extinguished.
Nevertheless, since the Na~I~D lines are almost certainly saturated, this value of $E(B-V)_{\rm host}$ should be taken as a lower limit only.
Combining  the Galactic reddening obtained from the infrared dust maps of  \citet{Schlegel98},
$E(B-V)_{\rm galactic}=$0.098, and the host galaxy value, we obtain a lower limit of the total reddening  of $E(B-V)_{tot} =0.44$ and $E(B-V)_{tot} =1.16$, respectively, bracketing the reddening obtained from the Balmer decrement.

Despite the relative proximity of NGC 3278, there are no direct distance measurements 
to this galaxy reported in the literature. 
We therefore must resort to the recession velocity and Hubble's law to compute the
distance to SN~2009bb.
In the remainder of this paper we assume a distance modulus of $\mu=33.01\pm0.15$ as given
in the NASA/IPA Extragalactic Database (NED). This value is obtained from
a  heliocentric velocity of 2961$\pm$37 
km~s$^{-1}$  \citep{Strauss92}, corrected for Virgo Infall and Great attractor \citep{Mould00}, and assuming $H_0=73.0\pm5$ \kms~Mpc$^{-1}$ \citep{Spergel07}.
Using the maximum light magnitudes reported in Table~4, and the previously mentioned values of reddening and distance, we compute absolute magnitudes for the various bands  
that are listed in Table~4.

Figure~\ref{fig3.1} shows the pseudo-bolometric light curve of SN~2009bb, together with  SN~1998bw, SN~2002ap, SN~2003jd and SN~2006aj, obtained by 
integrating the absolute flux in the $BVRI$ bands.
The peak luminosity of SN~2009bb is found to be close to that of SN~2006aj, and $\sim$60\% of that of SN~1998bw.

To estimate the $^{56}$Ni content and the ejected mass, we integrated the flux of SN~2009bb contained within the $u'BVRIJH$ bands. 
Also included was a $K$ band correction based on the SN~2002ap light curves  \citep{Yoshii03}. 
The resulting pseudo-bolometric light curve was then modeled  
making use of an analytical description 
for the peak of the light curve \citep{Arnett82}. The model assumes
spherical symmetry, homologous expansion, no mixing of $^{56}$Ni,
 radiation-pressure dominance, and
the applicability of the diffusion approximation for photons, which
restricts it to early phases ($<$ 30 days post-explosion) when the density is sufficiently high to make the ejecta optically thick.
 We also assume  constant opacity $k_{opt}$=~0.08~cm$^2$ g$^{-1}$ as computed for SN~1997ef by \citet{Mazzali00}. \citet{Chugai00} used a time variable opacity for modelling the SN~1998bw bolometric light curve which average value during the first 20 days after the explosion is $k_{opt}$=~0.07~cm$^2$ g$^{-1}$.  We assume the uncertainty on  visible opacity to be the difference of the previously mentioned values i.e. 0.01~cm$^2$ g$^{-1}$.  In \citet{Arnett82} model, the time-evolution of luminosity is given by

\begin{equation}
L(t)=M_{Ni}\epsilon_{Ni} e^{-x^2} \int_{0}^{x} 2ze^{-2xy+z^2}dz,
\label{equa3} 
\end{equation} 

\noindent where $\epsilon_{Ni}$ is the energy produced in 1 second by 1 gram of $^{56}$Ni, $x$=t/$\tau_m$ and  $y$=$\tau_m$/2$\tau_{Ni}$ with $\tau_{Ni}$ being the $e$-folding time of the $^{56}$Ni decay and $\tau_m$ effective diffusion time, which determines the width of the bolometric light curve.

If $\tau_m$ is expressed as a function of $k_{opt}$, the ejecta mass, $M_{ej}$, and the photospheric velocity, $v_{ph}$ at time of bolometric maximum,  the effective diffusion time  can be written \citep{Arnett82} as

\begin{equation}
\tau_m \sim \left(\frac{ k_{opt} M_{ej}} {v_{ph}}\right)^{1/2}.
\label{equa4} 
\end{equation}

In addition to  the M$_{\rm ej}$ and M$_{\rm Ni}$ parameters, there is also an additional free parameter,
T$_{\rm rise}$, which is the time interval between explosion and the peak of the bolometric light curve.  
Finally, as proposed by \citet{Valenti08}, we include the contribution of the $^{56}$Co decay in equation \ref{equa3}. 

In this analytic model, the photospheric velocity at the time of the maximum of the bolometric light curve is an input parameter. The velocity derived from the minimum of the \ion{Si}{2}~$\lambda$6355 feature is often used as a proxy for the  photospheric velocity. Nevertheless, several studies have found that in SN Ic the \ion{Si}{2}~$\lambda$6355 could be contaminated by other species like detached \ion{He}{1}~$\lambda$6678, \ion{Ne}{1}~$\lambda$6402, detached H$_{\alpha}$ or detached \ion{C}{2}~$\lambda$6580 \citep{Clocchiatti96,Branch06,Sauer06,Elmhamdi06}. Contamination from other ions  could shift the minimum of the \ion{Si}{2}~$\lambda$6355 feature in different direction at different epochs, biasing the estimation of the photospheric velocity. In spite of these shortcomings the \ion{Si}{2}~$\lambda$6355 feature is still the most easy feature to measure, but a correction must be applied to derive a velocity as close as possible to the real photospheric velocity at a given epoch.
To compute this correction we consider 
 SN~1998bw, SN~2002ap and SN~2006aj, for which the evolution of the photospheric velocity was computed using the Montecarlo code described in \citet{Mazzali01}.
Comparing the velocities measured from the minimum of the \ion{Si}{2} $\lambda$6355 absorption with those derived through the spectra modelling mentioned above,  a difference of $\sim$3000~\kms~ is found.
Assuming that this velocity shift also applies to SN~2009bb, we derive $v_{\rm ph}$ $=$ 15000 $\pm$ 1000~\kms~ at the time of the bolometric maximum light (JD=2454923.1 $\pm$ 0.7). The analytical model then yields M$_{\rm Ni}$ = 0.22 $\pm$ 0.06~M$_{\sun}$, M$_{\rm ej}$ = 4.1   $\pm$ 1.9~M$_{\sun}$ and T$_{\rm rise}$=11.3 $\pm$ 0.8 days. The uncertainties in the latter quantities were computed through Montecarlo simulations, taking into account the errors in the photospheric velocity, optical opacity, distance modulus, reddening and photometry.
 The resulting model is compared to the computed bolometric light curve in the insert of Figure~\ref{fig3.1}.

The ejecta mass and photospheric velocity allow us to estimate the kinetic energy associated with the 2009bb explosion by comparing the pseudo-bolometric light curve of SN~2009bb with that of SN~2002ap, for which detailed modelling have been  already carried out \citep{Mazzali02}.
To compute the kinetic energy, equation~\ref{equa4} can be written as follows:

\begin{equation}
\tau_m \propto  k_{opt}^{1/2} M^{3/4}_{ej} E_{kin}^{-1/4}.
\end{equation}

Around maximum the pseudo-bolometric light curves of SN~2002ap and SN~2009bb are similar (see the rescaled pseudo-bolometric light curve of SN~2002ap in the insert of Figure~\ref{fig3.1}). The decline rate of SN~2002ap is slightly slower than SN~2009bb. Around two weeks after $B$ maximum, which is the latest time we considered in the bolometric light curve fit, SN~2002ap reaches the same brightness of SN~2009bb $\sim$ 3 days later than the latter SN. Nevertheless,  the rise to maximum of SN~2002ap seems to be faster than that of SN~2009bb. Computing  the  rise time of the bolometric light curve  as difference between the explosion date and the epoch on which the bolometric light curve reach the maximum, we obtain a rise time of $\sim$ 13.5 and $\sim$ 10 days for SN~2009bb and SN~2002ap respectively. We therefore can assume $\tau_{m_{2009bb}}$ $\sim$ $\tau_{m_{2009ap}}$. Taking into account that for SN~2002ap $M_{ej}=2.5 \pm 0.5 M_{\sun}$ and $E_{kin}$ = 4 $\pm$ 1 $\times 10^{51}$erg \citep{Mazzali07}, for SN~2009bb we obtain  $E_{kin} =1.8 \pm 0.7  \times 10^{52}$erg.

\section{Evidence for a Central Engine}

Radio observations of SN~2009bb spanning $\Delta t \approx 17-100$
days after the explosion revealed evidence for an extraordinarily
luminous, long-wavelength counterpart, with a spectral luminosity $L_{\nu}$ at $\nu=4.9$ GHz of $\approx 5\times
10^{28}~\rm erg~s^{-1}$ \citep{Soderberg10}.  In comparison to the other
$\sim 150$ nearby SNe Ibc observed in the radio on a similar
timescale, SN~2009bb is a factor of 10 to $10^3$ more luminous,
rivaling the radio afterglow luminosities of the nearest GRB-SNe,
including GRB\,980425 associated with SN~1998bw \citep{Kulkarni98}.  The
multi-frequency radio observations were well described by a
self-absorbed synchrotron spectrum, produced as the blast-wave
shock-accelerated electrons in the local circumstellar medium.  Just
as in the case of SN~1998bw, the high brightness temperature of the
SN~2009bb radio counterpart required an unusually large emitting
region and, in turn, a trans-relativistic blast-wave velocity,
$v\approx 0.9c$.  Moreover, the high radio luminosity implied a
copious energy coupled to the fastest ejecta, $E\gtrsim 10^{49}$ erg.
\citet{Soderberg10} appealed to a model where this trans-relativistic
ejecta component was powered by a different mechanism, a ``central
engine'' (rapidly-rotating and accreting black hole or magnetar), as
is commonly assumed for long-duration GRBs \citep{MacFadyen01}.

This model can be tested using the bulk explosion parameters ($E_K$,
$M_{\rm ej}$) for SN~2009bb derived from our well-sampled optical
photometry and spectroscopy.  Analytic solutions for the distribution
of ejecta in Type Ibc SNe predict a steep coupling of energy and
velocity within the homologously-expanding material,

\begin{equation}
E(v)\approx 3.7\times 10^{47} \left(\frac{E_{\rm SN}}{10^{51}}\right)^{3.59} \left(\frac{M_{\rm ej}}{M_{\odot}}\right)^{-2.59} \left(\frac{v}{0.1c}\right)^{-5.18}~\rm erg
\label{eqn:energy}
\end{equation}

\noindent
\citep{Matzner99,Berger02}. Based on an analysis of the optical dataset
(see Section 3), we find bulk ejecta parameters of $E_K\approx 1.8\times
10^{52}$ erg and $M_{\rm ej}\approx 4.1~\rm M_{\odot}$.  Within the
homologous outflow, we therefore expect the ejecta traveling at
$v\gtrsim 0.9c$ to carry a kinetic energy of $E_K\approx 3.5\times
10^{45}$ erg.  This is {\it at least} three orders of magnitude below
the energy budget required to power the trans-relativistic radio
emitting ejecta.  We conclude that the luminous synchrotron emission
from SN\,2009bb cannot be explained in the framework of a homologous
SN explosion and another energy reservoir (i.e. a central engine) is
required to explain the distribution of ejecta within this  relativistic SN.

\section{Spectroscopy}

Optical spectra of SN~2009bb were obtained on thirteen epochs spanning phases between $-$2 and $+$285 days. Table~5 provides a log of these observations, and Figure~\ref{fig4.0} illustrates the spectral evolution.

\subsection{General evolution}

Comparing the  de-reddened spectra of SN~2009bb
to similar epoch spectra of  SN~1998bw, SN~1997ef, SN~2003jd and SN~2002ap, de-reddened using the color excess values mentioned in Section~3, one can note that at early phases SN~2009bb, SN~2003jd and SN~2006aj are characterized by a  bluer continuum than SN~1997ef  and SN~2002ap (see Figures~\ref{fig4.1} and \ref{fig4.2}). Around two week past maximum,  the continuum has become similar to that of all the other comparison SNe.

In terms of degree of broadening of the spectral features around  maximum, the spectra of SN~2009bb are similar to those of SN~2006aj  (see  Figure~\ref{fig4.1} and \ref{fig4.2}).
 Looking at the SN~2009bb spectra taken later than a week after $B$ maximum (see Figure~\ref{fig4.3},  Figure~\ref{fig4.4} and \ref{fig4.5}), the degree of blending is still high in SN~2009bb being  intermediate between SN~2003jd and SN~1998bw. This slow spectroscopic evolution in spite of the high expansion velocity is a further indication that SN~2009bb had a massive envelope. The expansion velocity of 2009bb decreases similarly to that of  SN~1998bw and  slower SN~2002ap (see Figure~\ref{fig4.8}).

\subsection{Detailed Comparison: the Helium Detection}

In the following, the evolution of some features present in the spectra of SN~2009bb are analyzed in more detail using  the supernova spectrum-synthesis code SYNOW \citep{Jeffery90,Branch03}. Particular effort is devoted to determining if helium is present in the ejecta of SN~2009bb, as this is very important for constraining the nature of its progenitor and
its evolutionary stage at the time of the explosion.
We explore qualitatively the presence of helium using  SYNOW, but it must be kept  in mind that full spectral modeling, including non-thermal effects is necessary to study quantitatively the evolution and strength of the He lines \citep{Lucy91}.
 
The clearest evidence of the presence of helium in the SN~2009bb ejecta is found in the first two spectra taken at $-$2 and $-$1 days (see Figure~\ref{fig4.1}).
A striking feature in these  two  spectra is the absorption around 5400~\AA. This line is not visible in the other SNe spectra, and we suggest that  it corresponds to \ion{He}{1} $\lambda$ 5876. Fitting the line profile with SYNOW, we obtain a good match for a velocity in the line forming region of $\sim$ 28000 \kms~  and 27500 \kms~ in the $-$2 and $-$1 days spectra, respectively. A similar velocity is also obtained  by fitting the absorption  around 4800~\AA~ with \ion{Fe}{2} $\lambda\lambda$4924, 5018, 5169 and also provides a good match with \ion{Si}{2} $\lambda$6355 for the absorption located at $\sim$5800 \AA.
In Figure~\ref{fig4.1}, the expected positions of the \ion{He}{1}~$\lambda$6678 and \ion{He}{1}~$\lambda$7065  absorption are also marked.
In both the $-$2 and $-$1 day spectra, features are present at this wavelength, but due to their weakness and to strong background contamination,  should not be over-interpreted.

In Figure~\ref{fig4.2} we show the SN~2009bb spectrum taken at $+$4 days. The velocity obtained from SYNOW  fits to \ion{Fe}{2}  $\lambda\lambda$4924, 5018, 5169  and \ion{Si}{2} $\lambda$6355 
is $\sim$ 18000 \kms.
SN~2009bb shows the closest resemblance to SN~2006aj, but differences are visible 
particularly around the \ion{Si}{2}~$\lambda$6355 absorption and  between 5000 and 6000 \AA, where  SN~2009bb clearly differs  compared to the other SNe.  Attributing this absorption  to  detached 
\ion{He}{1} at  $\sim$25000 \kms, we obtain a reasonable fit of the observed feature (see short dashed green line in Figure~\ref{fig4.6}).

The expected positions of the other \ion{He}{1} features  for the assumed 
detached velocity ($\sim$25000 \kms) are indicated in Figure~\ref{fig4.2}. Indeed, an absorption corresponding to 
\ion{He}{1}~$\lambda$6678 appears to be present and  helps to fit the boxy profile of the \ion{Si}{2} line 
(see Figure~\ref{fig4.6}). There is also a weak feature corresponding to  the expected position of 
\ion{He}{1}~$\lambda$7065, but contamination from \ion{Ne}{1}~$\lambda$6929 cannot be excluded.

Figure~\ref{fig4.3} shows the spectrum of SN~2009bb taken a week after $B$-band maximum.  Between 5000 and 5500 \AA, SN~2002ap and SN~2003jd show an absorption apparently due to \ion{Na}{1}. This feature is absent in the spectra of SN~2009bb, SN~1998bw and SN~1997ef, which is probably due to the broadness of the spectral features in the latter SNe, that tends to wash out weak features.
Nonetheless, the absence of a clear feature in the SN~2009bb makes less plausible the possibility that in the earlier SN~2009bb spectra the absorption around  5500 \AA~ is due to  \ion{Na}{1}.
As shown in Figure~\ref{fig4.6}, \ion{He}{1} with a detached velocity of 
$\sim$ 23000 \kms~ gives a good fit to the observed line profile between 
5000 and 5500 \AA.
Figure~\ref{fig4.3} also shows the spectrum of the broad lined Type~Ic SN~2007bg, for which the \ion{He}{1} lines (the positions of the \ion{He}{1}~$\lambda$7065, ~$\lambda$7235 lines are marked in the graph) appear to be even stronger than observed in the spectrum of SN~2009bb. 
Unfortunately, due to the higher expansion velocity of the absorption features
in SN~2009bb, the \ion{He}{1} $\lambda$7065 line 
falls in the gap between the CCDs, making its presence impossible to confirm.

Figure~\ref{fig4.4} shows a spectrum of SN~2009bb taken two weeks after $B$-band  maximum. 
The absorption feature around  5500 \AA~ is still deeper in SN~2009bb  than in the other two SNe.
For SN~2009bb, an equally good fit to this  feature is obtained with \ion{Na}{1} undetached ($\sim$ 14000 \kms), or with  \ion{He}{1} $\lambda$5876 detached at $\sim$ 18000 \kms. Again, the expected positions of the other prominent \ion{He}{1} lines are indicated in this Figure, but clear signs of these features cannot be seen in the SN~2009bb spectrum.

 Figure~\ref{fig4.5} displays the spectra taken on days $+$30 and $+$44. The emission redward of the 
 \ion{Na}{1} absorption remains stronger  in SN~2009bb compared to the
the other SNe plotted in this Figure, but the profile around the  \ion{Na}{1} absorption is now not so different suggesting a reduced strength of the blue wing of the  \ion{He}{1} $\lambda$5876  detached component. In both the $+$33 and $+$44 spectra of SN~2009bb, there also is no clear signature of the  \ion{He}{1}  $\lambda$6678 and $\lambda$7065 features. 
It should be noted that in the rather noisy spectrum of SN~2007bg, the helium lines, if present, are not as strong as expected. In the case of helium-rich core-collapse SNe (SNe~Ib), the \ion{He}{1} features are stronger at these later epochs than at maximum \citep[e.g.][]{Matheson01}. A possible explanation for the  lack of  strengthening of the 
\ion{He}{1} absorption features in the case of SN~2009bb (and perhaps also for SN~2007bg) is that, due to the higher expansion velocity of broad-lined SNe~Ic together with a less massive helium shell, the mass of helium in the line forming region at these later epochs is likely much lower than in normal SNe~Ib.

The evolution of the  \ion{He}{1} $\lambda$5876 velocity is summarized in the insert of  Figure~\ref{fig4.8}. Before maximum He is undetached from the main line forming region, while after maximum is detached. This agrees with the finding of \citet{Branch02} and \citet{Elmhamdi06} for Type Ib SNe. The velocity of the  detached He decrease with time. The latter also agree with what obtained by  \citet{Branch02}.

The fractional line depth of the \ion{O}{1} $\lambda$7774 line can, in principle, be used to infer the presence of helium using the formula given by \citet{Matheson01}

\begin{equation}
Fractional~line~depth~=~\frac{F_{cont}-F_{min}}{F_{cont}},
\label{equa9}
\end{equation}

\noindent where $F_{cont}$ and $F_{min}$ represent the values of the continuum and of the line minimum at the wavelength of the minimum, respectively.
The latter authors proposed that in SNe~Ib, due to the dilution of oxygen by helium, the  \ion{O}{1} $\lambda$7774 absorption should be  weaker than in SNe~Ic.
Indeed, they measured an average fractional line depth of   0.27 $\pm$ 0.11 and 0.38 $\pm$ 0.09 for SNe~Ib and   SNe~Ic, respectively.
\citet{Matheson01} also found that this  difference is stronger around maximum than at late time. Unfortunately, in broad-lined SNe~Ic, 
the depth of the \ion{O}{1} $\lambda$7774~\AA~ is difficult to measure at 
early epochs due to the widths of the other lines in this spectral region.
In Figure~\ref{fig4.7}, measurements at phases later than +20 days are plotted.
As for normal SN~Ibc the fractional line depth for broad-lined SNe~Ic increases with time.
Interestingly, the two broad-lined SNe~Ic which show evidence of helium lines, 
SN~2009bb and SN~2007bg, have an \ion{O}{1} $\lambda$7774 line depth that is
less than those that do not show any sign of helium.
Note that Figure~\ref{fig4.7} does not include SN~1998bw since, even at +50 days, \ion{O}{1} $\lambda$7774 is still strongly blended with the \ion{Ca}{2} near-infrared triplet.

\subsection{Nebular Spectrum}

In Figure~\ref{fig4.9} we show a comparison between the nebular spectra  of SN~2009bb, SN~1997ef, SN~1998bw, SN~2002ap and  SN~2006aj.
To study the line profiles of SN~2009bb, the strong emission lines from the underlying \ion{H}{2} region must be removed. This was achieved by subtracting the spectrum of the bright \ion{H}{2} region located $\sim$~0.3$^{''}$ North,  $\sim$~1.4$^{''}$ East of
the SN position. The latter spectrum was acquired with the same spectrograph, making more accurate the spectra matching procedure. The spectrum of the bright \ion{H}{2} region was rescaled to match the line fluxes of the \ion{H}{2} region underlying the SN. Note that the line ratios for the two \ion{H}{2} regions are similar to within 10\%. This, together with the very similar Balmer decrement measured in the two \ion{H}{2} regions (see section 3), gives confidence that the subtraction procedure has not significantly altered  the SN spectrum.

The nebular spectrum of SN 2009bb is rather noisy, but several strong emission
features can be recognized.  The emission in the blue, near 4500\AA, should be
dominated by [\ion{Mg}{1}] $\lambda$4570. The broad complex near 5200\AA~ is the result of a
number of [\ion{Fe}{2}] emission lines. Near 5900\AA, \ion{Na}{1}~D emission is possibly
detected. The two strongest lines in the spectrum are [\ion{O}{1}] $\lambda$$\lambda$6300, 6363 and \ion{Ca}{2}] $\lambda$$\lambda$7291, 7324.

It is immediately clear that the calcium to oxygen ratio is much larger in SN~2009bb than in all the other comparison SNe. 
It is also evident, when compared to the spectra of SNe 2003jd
and 1998bw, that the profile of the [\ion{O}{1}] $\lambda$$\lambda$6300, 6354 emission in SN 2009bb is more similar to that of SN~1998bw than that of SN~2003jd  (see Figure~\ref{fig4.10}).
Based on spectral modelling, \citet{Mazzali05} claimed that a double peaked  [\ion{O}{1}] $\lambda\lambda$6300, 6354 profile as observed in SN~2003jd is produced by an aspherical expansion, but viewed off axis. Assuming that this interpretation is correct [but see also \citet{Maeda08,Maurer10}], in the case of SN~2009bb, the material ejected  must have been oriented close to the line of sight as in the case of SN~1998bw.

The nebular spectrum of SN~2009bb can be modelled in order to derive some physical properties
of the SN. Nebular spectroscopy reveals the conditions
of the innermost part of the SN ejecta, which are not visible when the SN is
bright because they lie too deep in regions that are optically thick. 
At late phases, on the other hand, the inner parts of the SN are transparent, and
behave like a nebula.

The code we used computes the energy deposition from the decay of $^{56}$Ni to $^{56}$Co to $^{56}$Fe. These decays emit $\gamma$-rays and high-energy positrons,
which can deposit their energy in the SN ejecta and heat the gas by collision excitation.
Cooling then takes place via line emission. Our code was  described by \citet{Mazzali01}, and is based on the description of \citet{Axelrod80}.

We model the spectrum of SN 2009bb using the simple, one-zone version of the
nebular code. This assumes that the SN ejecta are spherical, and bound by an
outer velocity, which reflects not so much the extent of the ejecta in radius or
velocity (the two are equivalent in the homologously expanding SN ejecta), but
rather that of the region which is effectively heated and can therefore emit
lines. This may be the result of decreasing density, distribution of $^{56}$Ni, or
both. 

Our model requires an outer velocity of 5500 \kms. This velocity appears to be adequate to fit all lines.
This could be taken as an argument that SN~2009bb was not significantly aspherical. 
In models where more energy is released in a polar direction, leading to the synthesis of $^{56}$Ni and possibly to the production of a GRB, the [\ion{Fe}{2}] lines are broader than the [\ion{O}{1}] line if the
event is viewed close to the polar axis \citep{Maeda03}. Otherwise, the
[\ion{Fe}{2}] lines are narrower than the [\ion{O}{1}] line, which shows a characteristic
double-peaked profile, as in SN~2003jd.
Nevertheless, the strongest evidence for asphericity in SN~1998bw was seen in nebular
spectra obtained ~200 days after maximum \citep{Mazzali01},
while the difference between the  iron and oxygen lines width was smaller at epochs of ~340 days,
which is closer to the epoch of the only nebular spectrum available for SN~2009bb.
Therefore, while from this spectrum only we do not infer major asphericities
in SN~2009bb, we cannot rule out that the signature of asphericity might have been seen in
earlier nebular data. 

The $^{56}$Ni mass which is required to reproduce the spectrum, assuming a distance
modulus of 33.0 and a reddening $E(B-V)=0.58$, is 0.25 M$_{\sun}$. This is in good agreement with the estimate of 0.22$\pm$0.06 M$_{\sun}$ found from  modelling the pseudo-bolometric light curve.
 The mass of the elements used to generate the synthetic spectra are summarized in Table~6.

\section{Discussion and conclusions}

We have presented UV, optical and near-infrared photometry, and optical spectroscopy
of the broad-lined Type~Ic SN~2009bb.
Around  maximum, the spectra of SN~2009bb were similar to those of SN~2003jd and SN~2006aj. The expansion velocities of  SN~2009bb are similar to these of SN~2006aj, but higher than those of SN~2003jd.
At phases later than +15 days, the broadness of the spectral features in SN~2009bb was intermediate between SN~1997ef and SN~1998bw.
The slow spectroscopic evolution indicates that SN~2009bb had a massive envelope, and this
is  confirmed by the pseudo-bolometric light curve analysis. With a simple model that includes both the $^{56}$Ni and  $^{56}$Co decay,  and making use of  analytic equations presented in \citet{Arnett82}, we obtain a  $^{56}$Ni and  ejected  mass of  0.22$\pm$0.06 M$_{\sun}$ and 4.1 $\pm$1.9 M$_{\sun}$, respectively. The resulting kinetic energy is $18 \pm 7 $ Foe.
A similar $^{56}$Ni mass is obtained from the nebular spectrum modelling $M_{^{56}Ni} = 0.25 M_{\sun}$.

An absorption feature is identified in the $-2$ and $-1$ day spectra  around 5400~\AA~ that we  attribute to \ion{He}{1} $\lambda$5876  forming at the photosphere.
\ion{He}{1} appears to be detached in the later spectra, with the absorption getting weaker with time.
The latter evolution is odd compared to what is normally observed in SNe~Ib where helium lines  become stronger after maximum.
The weakening of the helium absorption could be due to a combination of  the fast expansion characteristic of broad-lined SNe and a less massive helium shell than in ``normal'' SN~Ib.

The comparison between the kinetic energy carried away by the homologous expanding ejecta at $v\gtrsim 0.9c$ and the energy inferred by the radio observations  points to the need for an extra source of energy  provided by a central engine.

The modelling of the SN~2009bb nebular spectrum  indicates that  iron and oxygen lines can be reproduced with a synthetic spectrum where both elements have similar expansion velocities. This suggests that  SN~2009bb ejecta structure was not significantly aspherical. Nonetheless, considering the iron and oxygen lines evolution of SN~1998bw, we cannot exclude that the signature of asphericity could have been detected in a SN~2009bb earlier nebular spectrum. 
The analysis of the   [\ion{O}{1}] $\lambda\lambda$6300, 6354 reveals that, if the explosion was indeed aspherical, the orientation should have been similar to that of SN~1998bw.
Therefore, if a GRB was produced during the SN2009bb explosion, it was below the threshold of the current generation of $\gamma$-ray instruments.  
Without a detected GRB counterpart, SN~2009bb represents the
first engine-driven, relativistic supernova ever discovered by its
optical emission alone. 
The presence of a relatively massive helium layer may have played a role on the failed GRB detection, but a quantitative description of the helium shell is necessary to verify this hypothesis.

\acknowledgments
A special thanks to the Gemini-South staff for their efforts in obtaining data for our spectroscopic program.
G.P. acknowledges support by the Proyecto FONDECYT 11090421 and  from Comit\'e Mixto ESO-Gobierno de Chile. M.H. ackowledges support from FONDECYT through grant 1060808.
G.P., M.H. and J.M. acknowledge support from the Millennium Center for Supernova Science through grant P06-045-F funded by ``Programa Bicentenario de Ciencia y Tecnolog\'ia de CONICYT'', ``Programa Iniciativa Cient\'ifica Milenio de MIDEPLAN''. 
G.P., M.H. and J.M. acknowledge partial support from Centro de Astrof\'isica FONDAP 15010003 and by Fondecyt through grant 1060808 from the Center of
Excellence in Astrophysics and Associated Technologies (PFB 06).
The Dark Cosmology Centre is funded by the Danish NSF.
This material is based upon work supported by the National Science
Foundation (NSF) under grant AST--0306969. 
This paper is based on observations obtained at the Gemini Observatory, Cerro Pachon, Chile (Gemini Programs GS-2009A-Q-17 and  GS-2009A-Q-43).
We have made use of the NASA/IPAC Extragalactic
Database (NED) which is operated by the Jet Propulsion Laboratory,
California Institute of Technology, under contract with the National
Aeronautics and Space Administration. 






\appendix 
\section{Observations and data reduction}

Imaging of SN~2009bb was acquired  with five different instruments in the optical, one in the near-infrared,  and one in the ultraviolet. A summary of the characteristics of the facilities used are listed below.

\subsection{Photometry}

\begin{itemize}

\item The SWIFT space telescope is equipped with the UVOT  camera that has 256$\times$256 physical pixels, but after  processing the final combined images has 2048$\times$2048 pixels with a pixel scale of 0\farcs5. SN~2009bb was observed with the $uvw1$ and $u$ filters.

\item The PROMPT 3 telescope at CTIO is equipped with a CCD camera Alta U47UV E2V CCD47-10 (1024$\times$1024, pixel scale = 0\farcs6 per pixel). SN~2009bb was observed with both a standard Johnson-Kron-Cousins $B$-band and a Sloan $g'$ filters.

\item PROMPT 5 is equipped with a CCD camera Alta U47 E2V CCD47-10 (1024$\times$1024, pixel scale = 0\farcs6 per pixel). SN~2009bb was observed with both Johnson-Kron-Cousins $VRI$ and Sloan  $r'i'z'$ filters.

\item The 1.0~m Henrietta Swope telescope at LCO is
 equipped with a SITe3 CCD camera (2048$\times$3150, pixel scale = 0\farcs435  per pixel).  SN~2009bb was observed with both  Johnson-Kron-Cousins  $BV$ and  Sloan $u'g'r'i'$  filters.

\item The Swope telescope was also used to obtain near-infrared $YJH$ imaging with the RetroCam camera (HAWAII-1 1024$\times$1024, pixel scale = 0\farcs54)

\end{itemize}

All optical images were reduced following standard procedures
including bias, dark (when appropriate), and flat-field corrections. 
Photometry of the supernova was computed relative to a 
 local photometric sequence in the field of NGC~3278. The photometric sequence itself was calibrated
 to the standard Johnson Kron-Cousins and Sloan photometric systems using observations of   
 \citet{Landolt92} and \citet{Smith02}  photometric standard stars obtained over the course of seven photometric nights.
Our adopted $BVRI$ and $u'g'r'i'z'$ photometry of the local sequence is reported in 
Table~7 and Table~8, respectively.  The listed values 
were computed via a weighted average of the measurements made for 
each of the seven calibration nights.

 Given that the background of SN~2009bb was quite complex due to the presence of a bright \ion{H}{2} region, it was necessary to 
 apply template subtractions to all of the optical images. 
Three template images for each filter   were acquired with the PROMPT telescopes between 2010 May 25--30, or $\sim$ 420 days after $B$ maximum brightness.  To estimate the SN residual flux in these templates, we rescaled the $BVRI$ light curves of SN~1998bw in order to obtain the best match with SN~2009bb between +30 and +50 days. The brightness decline of SN~1998bw is slower than that of SN~2009bb (see Figure~\ref{fig2.1}), therefore a rescaling computed around maximum light should grossly overestimate the late-time flux of SN~2009bb.
Interpolating the rescaled light curves at the epoch on which the SN~2009bb templates were acquired, we estimate the  SN~2009bb brightness to be: $B$~$\sim$~24.6 mag, $V$~$\sim$~24.3 mag, $R$~$\sim$~23.3 mag and $I$~$\sim$~23.1 mag.
Assuming zero flux in the templates, these values imply an underestimation of the  SN~2009bb real flux in all the bands of less than 0.5\% around maximum, and less than 2\% on the last measurement published in this paper.
These values make us confident that the analysis reported  here is  not affected by systematic errors related to template subtraction.
Using the SN~2009bb spectra acquired on 2010 Jan. 9 UT and the standard bands of the Johnson-Kron-Cousins and Sloan photometric systems,  we can estimate the previously mentioned systematic errors also for the $g'r'i'z'$ PROMPT light curves. As expected, also in this case the flux underestimation turns out to be negligible.
Taking into account that the decline rate of SN~1998bw is slower than SN~2009bb the previously reported SN~2009bb late phase magnitudes should be regarded as an upper limit of its real brightness. Nevertheless, possible late time flattening on the light curve cannot be excluded a priori.
In order to obtain a direct estimate of the SN~2009bb residual flux in the template images, we examined the IMACS $g'r'i'$ images taken on 28 Dec. 2009 UT. We measured the flux in an aperture centered in the SN position and of a radius equal to 3 times the sigma of the Gaussian profile of the stellar objects present in the image.
Through these measurements, we estimate upper limits of 20.9 mag, 20.5 mag and 20.1 mag for the $g'r'i'$ filters respectively. We note that most of the flux measured in the IMACS images is due to the  galaxy emission, not from the SN itself.
To prove this, we computed synthetic magnitudes on the  SN~2009bb nebular spectrum acquired on 9 Jan. 2010 UT  once this was cleaned from the \ion{H}{2}  region and stellar emissions. We obtained: $B$=24.0, $V$=23.9, $R$=23.0,  $I$=23.1, $g'$=23.84, $r'$=23.4, $i'$=23.4 and $z'$=23.8 mag.

In the case of the Swope telescope, the template images were acquired on 20 Nov 2009 UT. The SN flux in the $u'$ and $B$ bands, was negligible, but this was not the case in the $Vg'r'i'$ images. In order to  reliably calibrate the magnitudes in the light curve tail, we estimated the residual flux in the template images matching the PROMPT and Swope light curves around maximum. 

We next proceeded to compute $BVRIg'r'i'$ photometry of SN~2009bb 
in the standard Johnson and Sloan systems using the S-correction technique \citep{Stritzinger02}, but following the prescription of \citet{Pignata08}.
Since the $u'$ band is not fully covered by any of our SN~2009bb spectra, the SN  magnitudes were calibrated in the standard photometric system using the following  equation:

\begin{equation}
u^{'} = u + 0.044 \times (u^{'}-g^{'}) + ZP_{u^{'}}
\label{equal2}
\end{equation}

\noindent where $u^{'}$  is the magnitude in the standard photometric system and $u$ is the magnitude in the  instrumental one, while $ZP$ refers to the photometric  zero point.   
  
 The near-infrared images were obtained using a standard jitter technique.
 To reduce the data cubes, we made use  of  a pipeline developed by the CSP 
 \citep[for details see][]{Contreras10}. 
 The pipeline applies to each image dark, flat-field, and detector linearity corrections.
These altered images are geometrically aligned and then stacked to 
obtain a master image from which photometry was computed.
  
 Near-infrared photometry of the SN was also obtained relative to a local
 sequence of stars calibrated against \citet{Persson98} standard stars obtained during  three  photometric nights. The $YJH$ photmetry of the local sequence stars are reported in Table~9.
Comparing  our $J$ and $H$ magnitudes with these of the 2MASS catalog, we obtain a difference of 0.03 $\pm$ 0.03 mag and 0.02 $\pm$ 0.05 mag, respectively.
 
The near-infrared photometry instrumental magnitudes were computed  through the template subtraction technique. As the photometry of the SN is computed differentially with respect to the 
local sequence, we obtained final magnitudes using the following equations:
  

\begin{equation}
Y = y + ZP_{Y}
\label{equa6.1}
\end{equation}

\begin{equation}
J = j + ZP_{J}
\label{equa6.2}
\end{equation}

\begin{equation}
H = h + ZP_{H}.
\label{equa6.3}
\end{equation}

\noindent Here $YJH$ are the magnitudes in the standard photometric system, $yjh$ are the magnitudes in the natural photometric system and $ZP_{Y}$, $ZP_{j}$, $ZP_{h}$ are the zero points.

The UVOT images were reduced using the customized instrument pipeline that, in addition to preparing the frames for photometry, computes an astrometric solution for them.
Aperture photometry was performed on the RA, DEC position of SN~2009bb with the optimum  aperture radius of 10''  suggested by \citet{Poole08}.
To ensure that the aperture includes a constant amount of light, the stability of the PSF size was checked in all the combined images. The $r.m.s.$ of the FWHM variation was 0.2 pixels for both the $uvw1$ and $u$ UVOT filters.

\subsection{Spectroscopy}

Optical spectroscopy of SN~2009bb was obtained on thirteen epochs with 
the du Pont (+ WFCCD \& B\&C) and Magellan (+IMACS \& LDSS3) telescopes at LCO, and the Gemini South (+ GMOS) telescope. 
 Spectra were reduced in a standard manner using IRAF
scripts based on tasks contained within the IRAF {\tt noao.twodspec} and {\tt gemini.gmos} packages.
Optimal extraction was obtained by weighting the signal
according to the intensity profile along the slit. Sky subtraction was
generally carried out by fitting a low order polynomial to both sides of the extracted SN
spectrum, and wavelength solutions were determined from exposures of arc lamps.
The wavelength calibration was checked against bright night-sky
emission lines. Flux calibration was performed by means of
spectrophotometric standard stars \citep{Hamuy92,Hamuy94} and checked
against the photometry.
When discrepancies occurred, the flux of the
spectrum was scaled to match the broad-band photometry.




\clearpage



\begin{figure}
\epsscale{.80}
\plotone{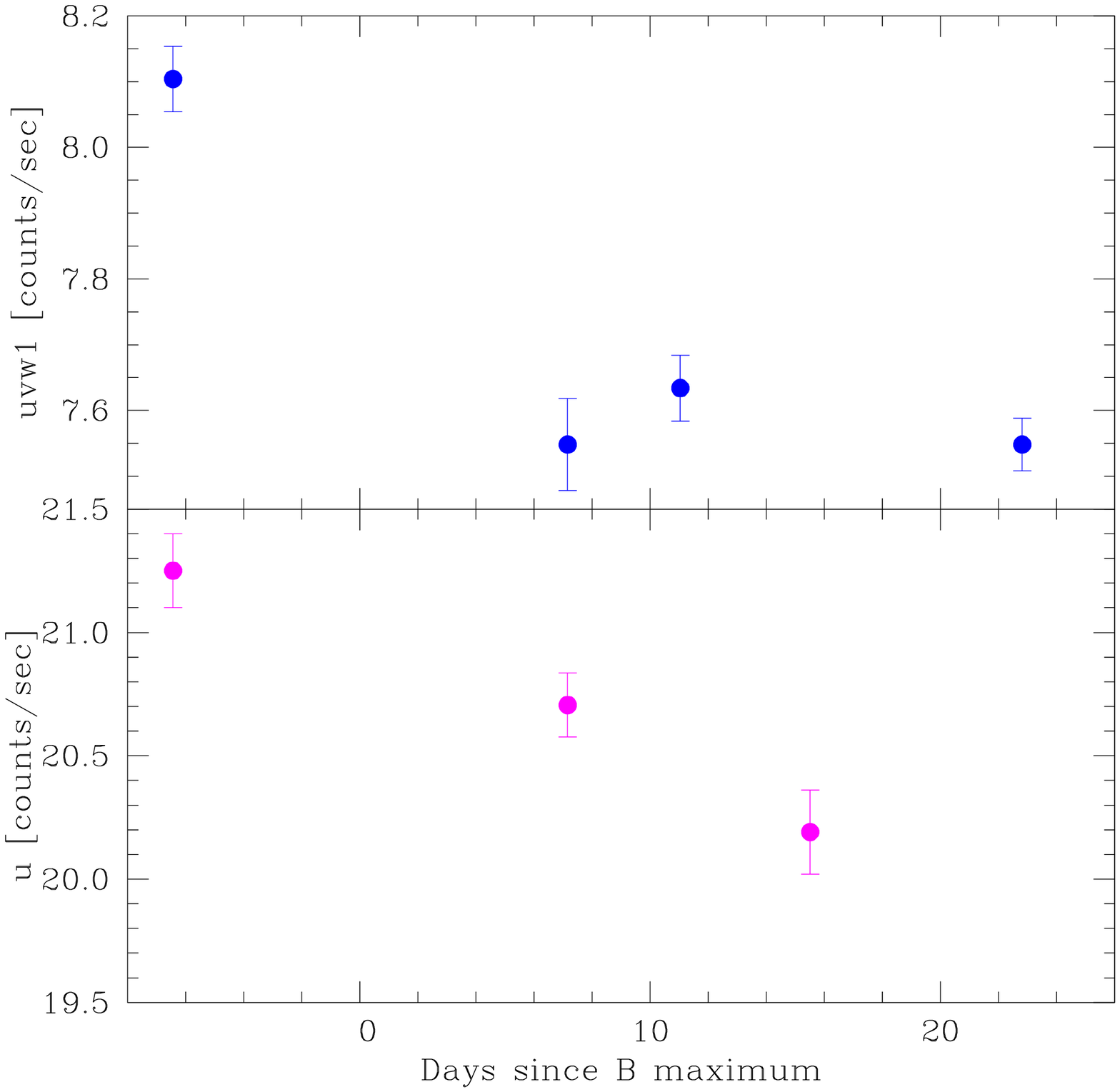}
\caption{Counts rate evolution at the position of SN~2009bb in the  $uvw1$ (top panel) and $u$ (bottom panel) bands.}
\label{fig2.0}
\end{figure}

\begin{figure}
\epsscale{.80}
\plotone{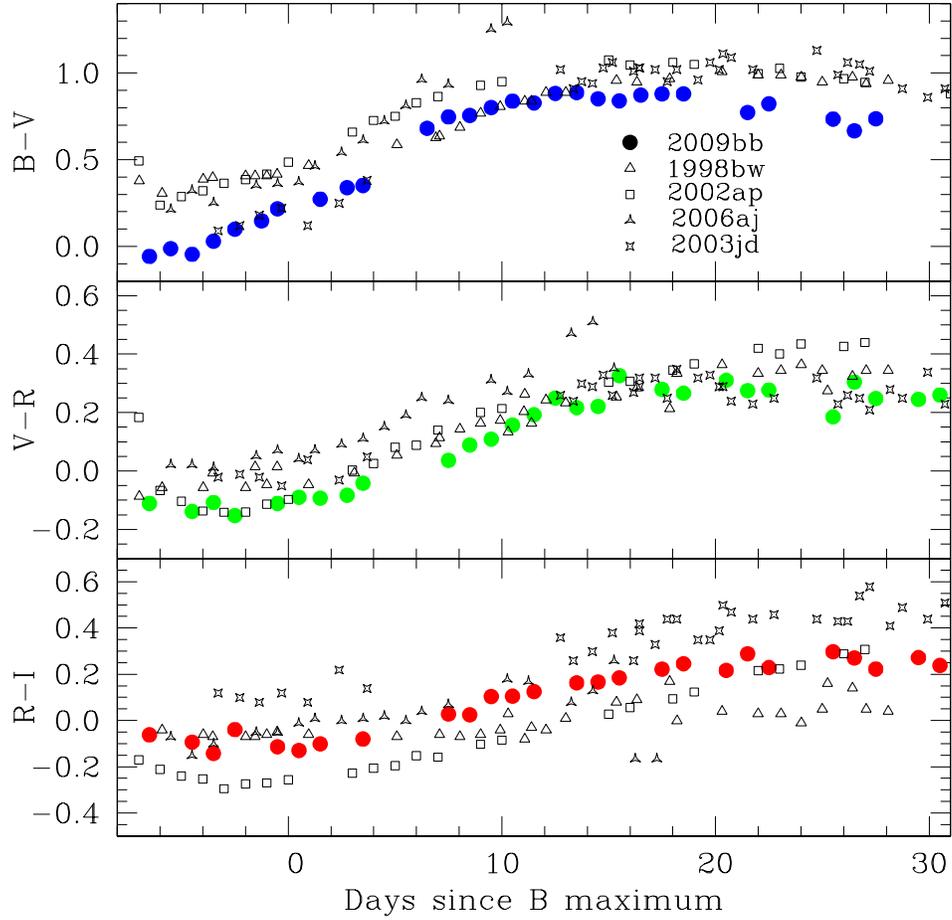}
\caption{De-reddened $(B-V)_0$, $(V-R)_0$ and $(R-I)_0$ color curves of SN~2009bb. For
comparison, the color curves of SN~1998bw, SN~2002ap, SN~2006aj and SN~2003jd are also shown. The bibliographic sources  are the same as those given in Figure~\ref{fig2.1}.}
\label{fig2.4}
\end{figure}

\begin{figure}
\epsscale{.80}
\plotone{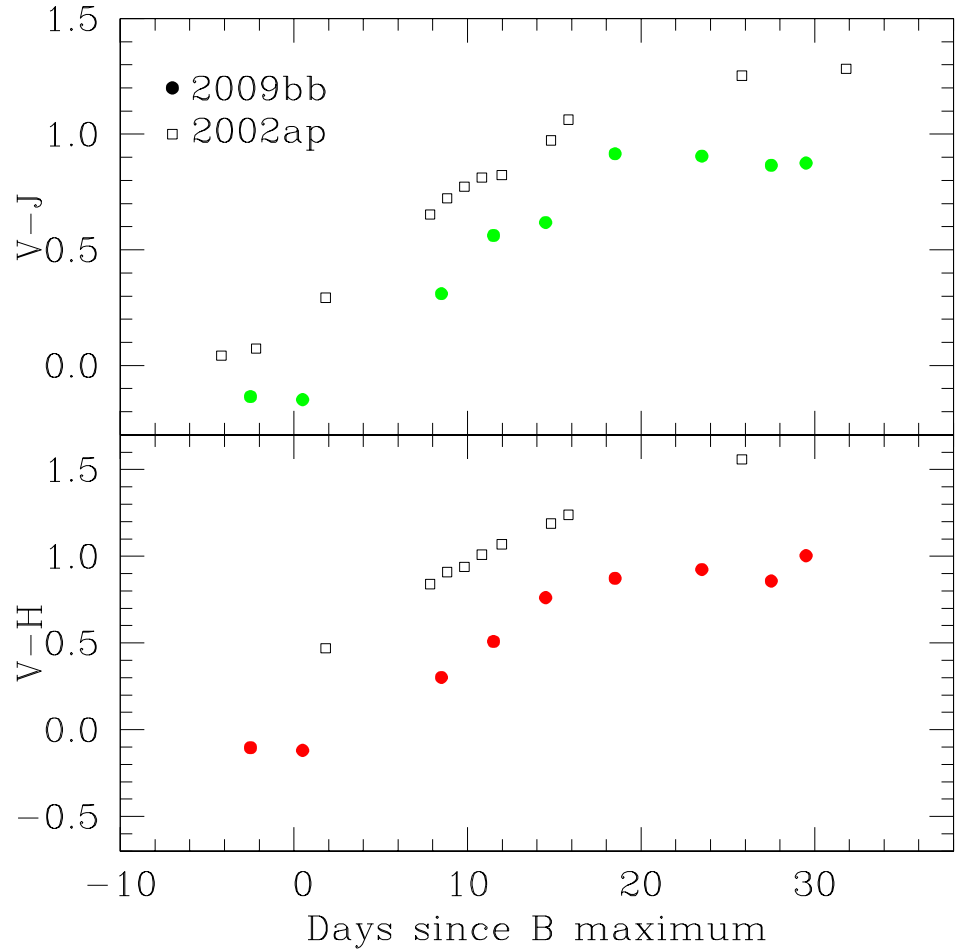}
\caption{De-reddened $(V-J)_0$ and $(V-H)_0$ color curves of SN~2009bb. For
comparison, the color curves of SN~2002ap are also shown. The bibliographic sources are the same as those given in Figure~\ref{fig2.3}.}
\label{fig2.5}
\end{figure}

\begin{figure}
\includegraphics[scale=.60,angle=90]{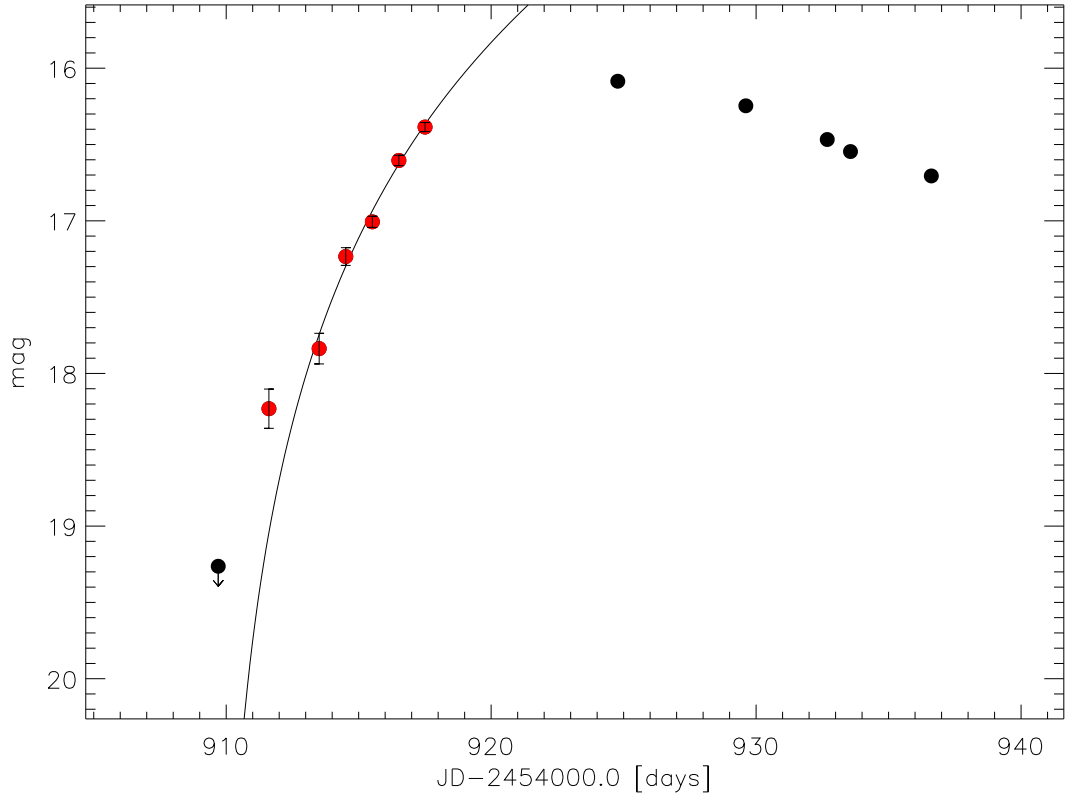}
\caption{The black filled circles are the unfiltered magnitudes of SN~2009bb calibrated to the $V$ band magnitude scale. The red filled circles are the measurements considered in the fit of the fireball model (solid line) used to estimate the SN explosion date. The first point indicates the upper-limit on the image taken immediately previous to the SN detection.}
\label{fig2.6}
\end{figure}

\begin{figure}
\epsscale{.80}
\plotone{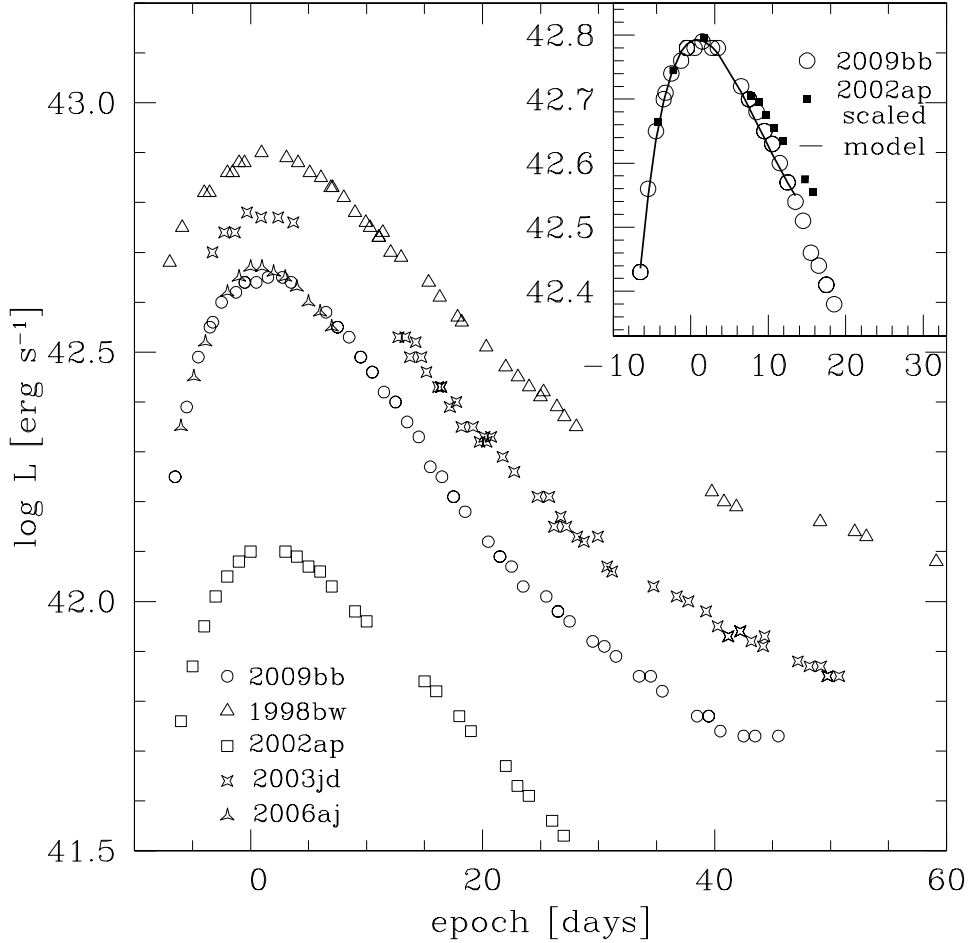}
\caption{In the main plot, the pseudo-bolometric light curve of SN~2009bb obtained by combining the flux in the $BVRI$ bands is compared with those of SN~1998bw, SN~2002ap, SN~2003jd and SN~2006aj calculated in a similar way. In the inset plot, the pseudo-bolometric light curve of SN~2009bb obtained combining the flux of the $u'BVRIJH$ bands with a $K$ band correction is shown together with the similar light curve of SN~2002ap rescaled to mach SN~2009bb around maximum.
The  model (solid line)  used to estimate the $^{56}$Ni and ejecta mass for SN~2009bb is also plotted.}
\label{fig3.1}
\end{figure}

\begin{figure}
\plotone{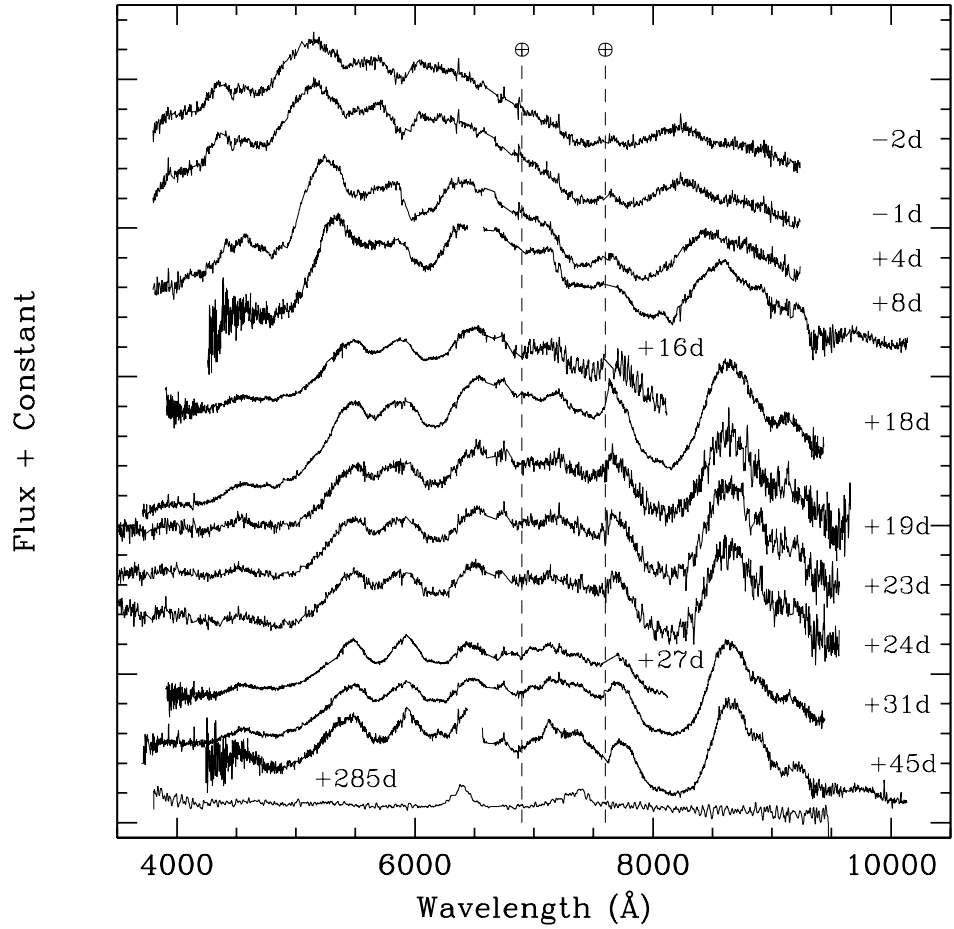}
\caption{Optical spectral evolution of SN~2009bb. For presentation
purposes the host galaxy emission lines have been removed and the spectra arbitrarily shifted. The $\earth$ symbol
shows the position of the main telluric features. The spectra are
labeled with the epoch in days past $B$ maximum.}
\label{fig4.0}
\end{figure}

\begin{figure}
\plotone{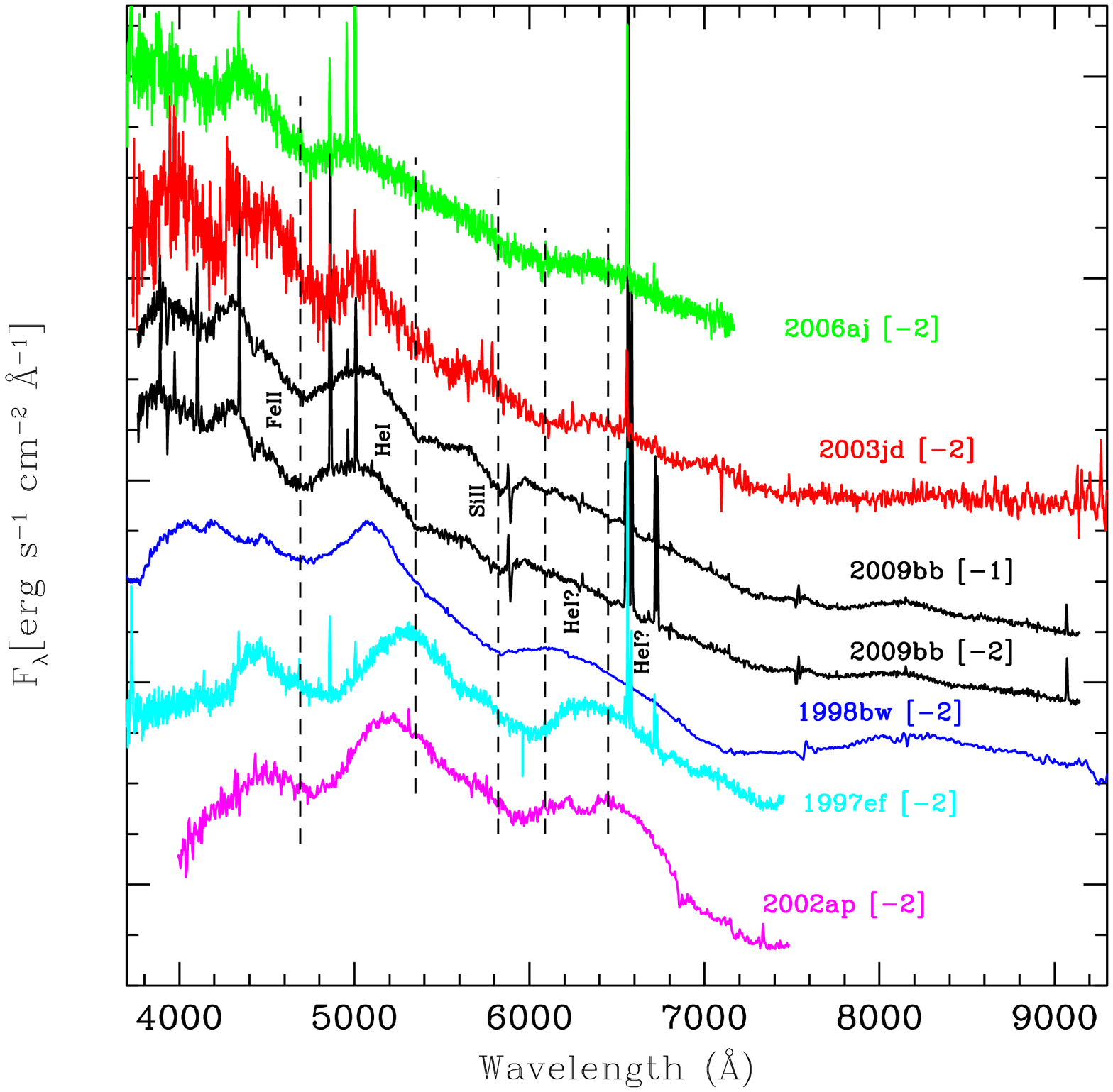}
\caption{First two spectra of SN~2009bb taken before maximum light. The phase with respect to $B$ maximum is noted in square brackets. Coeval
spectra of SN~1998bw \citep{Patat01}, SN1997ef \citep{Matheson01},
SN~2003jd \citep{Valenti08}, SN~2002ap \citep{Gal-Yam02} and SN~2006aj \citep{Modjaz06} are
shown for comparison. The spectra have been corrected for
reddening and redshift. The elements responsible for some absorption features in the SN~2009bb spectra are also labeled.}
\label{fig4.1}
\end{figure}

\begin{figure}
\plotone{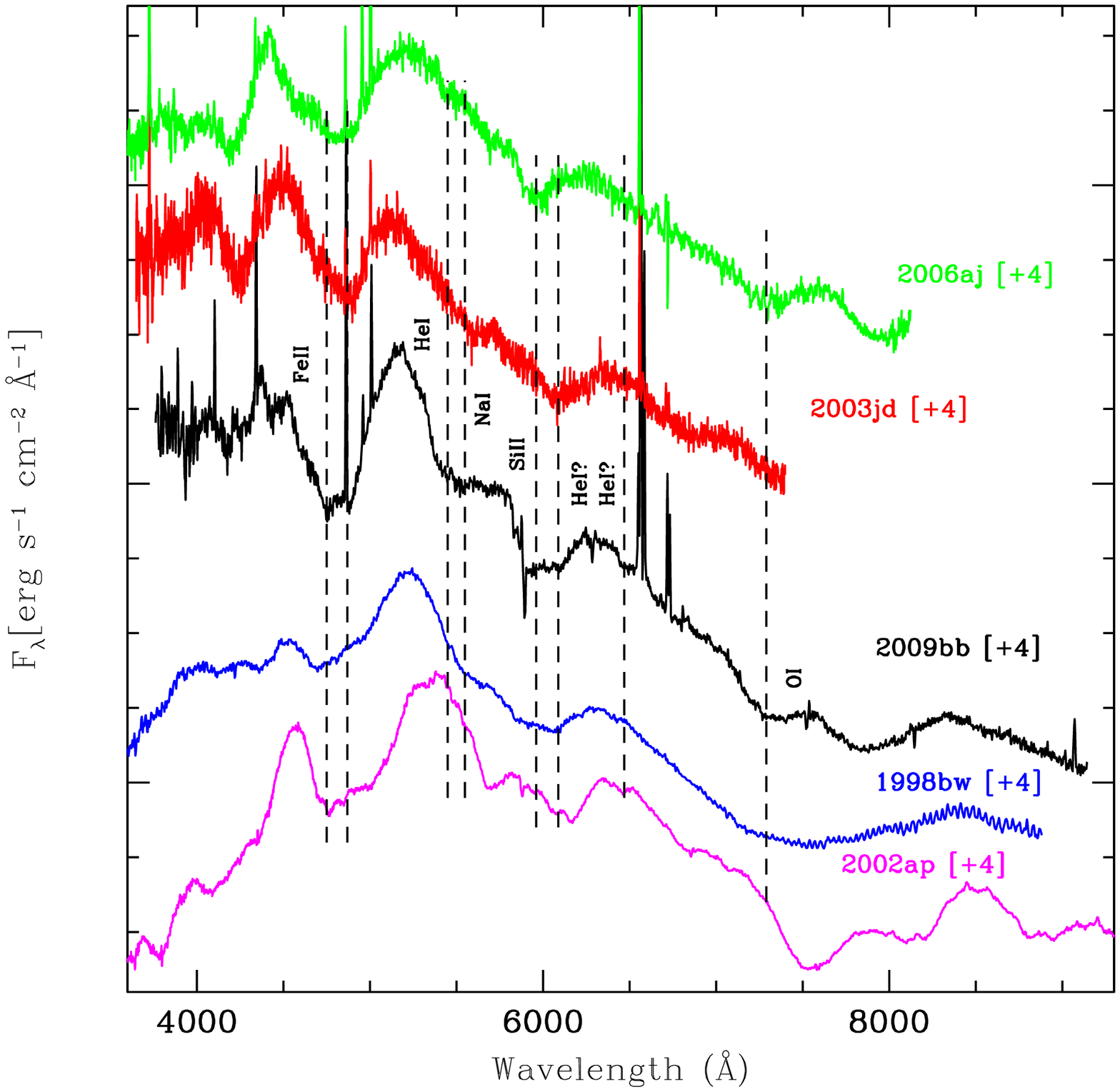}
\caption{Spectrum of SN~2009bb taken 4 days after $B$-band maximum. The bibliographic sources for the spectra of the  other SNe shown for comparison are the same as those in Figure~\ref{fig4.1}.}
\label{fig4.2}
\end{figure}

\begin{figure}
\plotone{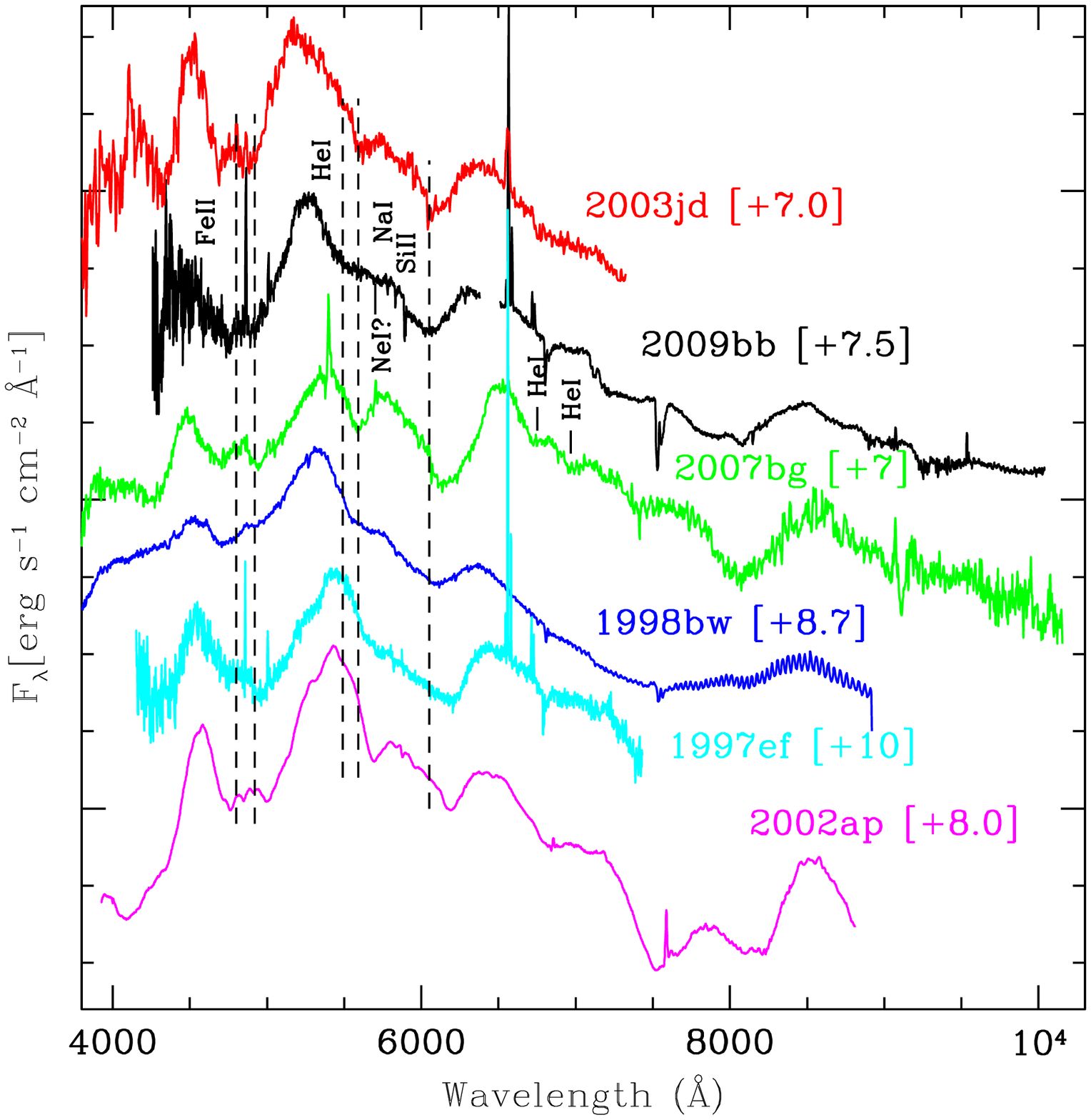}
\caption{Spectrum of SN~2009bb taken 7.5 days after $B$-band maximum. A spectrum of SN~2007bg \citep{Young10} is also reported. The bibliographic sources for the spectra of the  other SNe shown for comparison are the same as those in Fig~\ref{fig4.1}.}
\label{fig4.3}
\end{figure}

\begin{figure}
\plotone{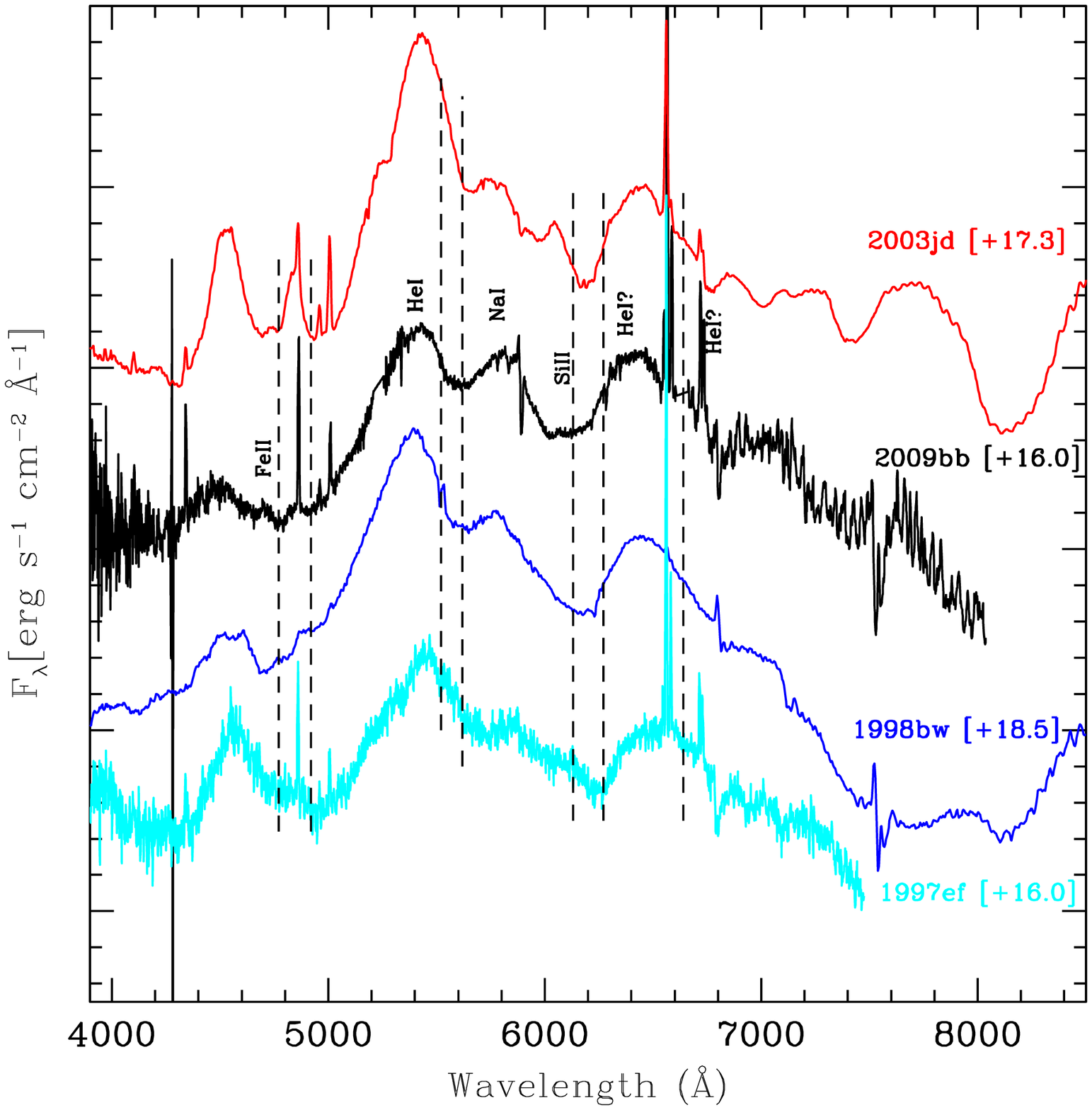}
\caption{Spectrum of SN~2009bb taken 16 days after $B$-band maximum. The bibliographic sources for the spectra of the  other SNe shown for comparison are the same as those in Figure~\ref{fig4.1}. }
\label{fig4.4}
\end{figure}

\begin{figure}
\plotone{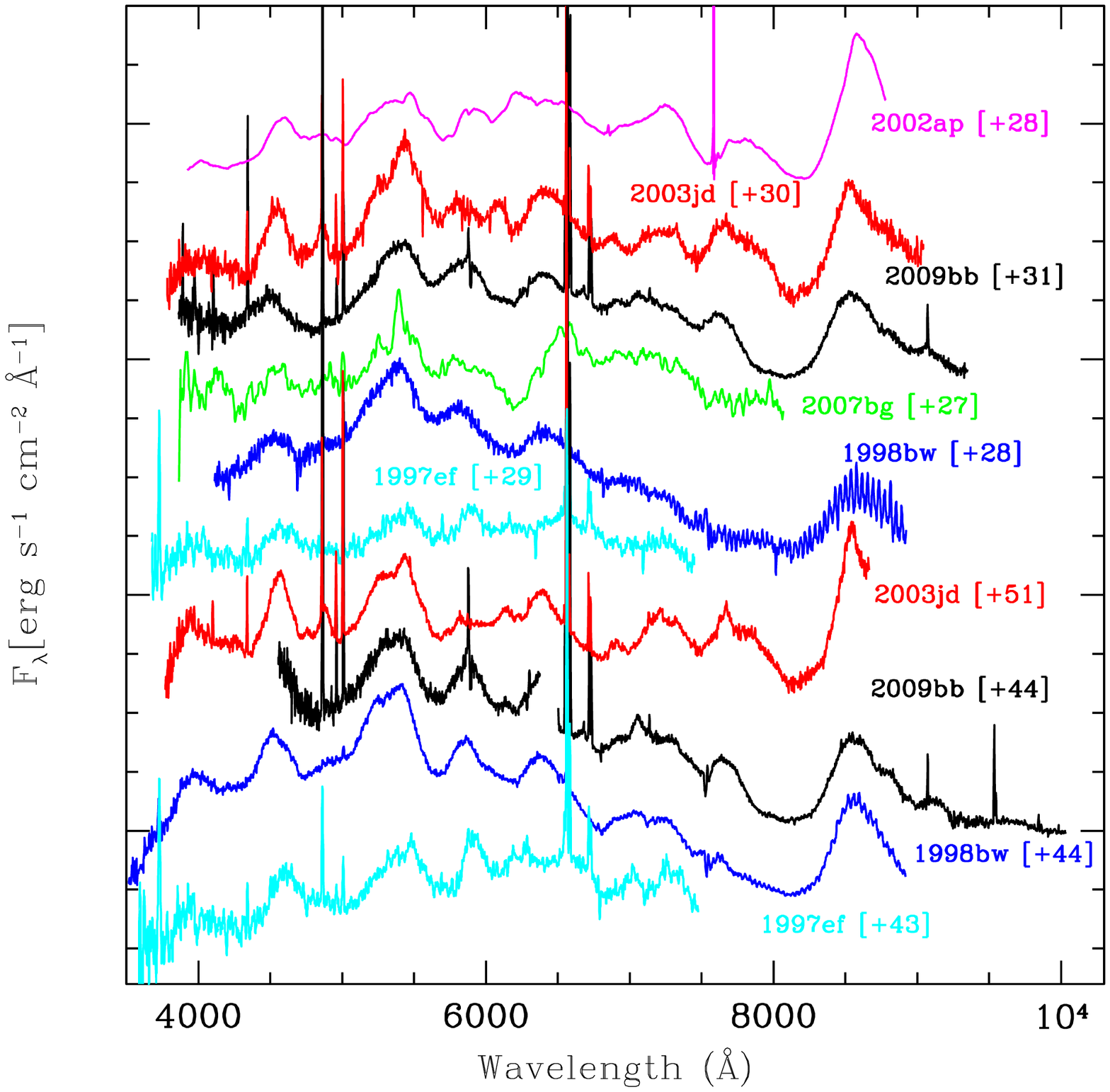}
\caption{Spectra of SN~2009bb taken 30 and 44 days after $B$ band maximum. The bibliographic sources for the spectra of the  other SNe shown for comparison are the same as those in Figure~\ref{fig4.1} and Figure~\ref{fig4.3}.}
\label{fig4.5}
\end{figure}

\begin{figure}
\plotone{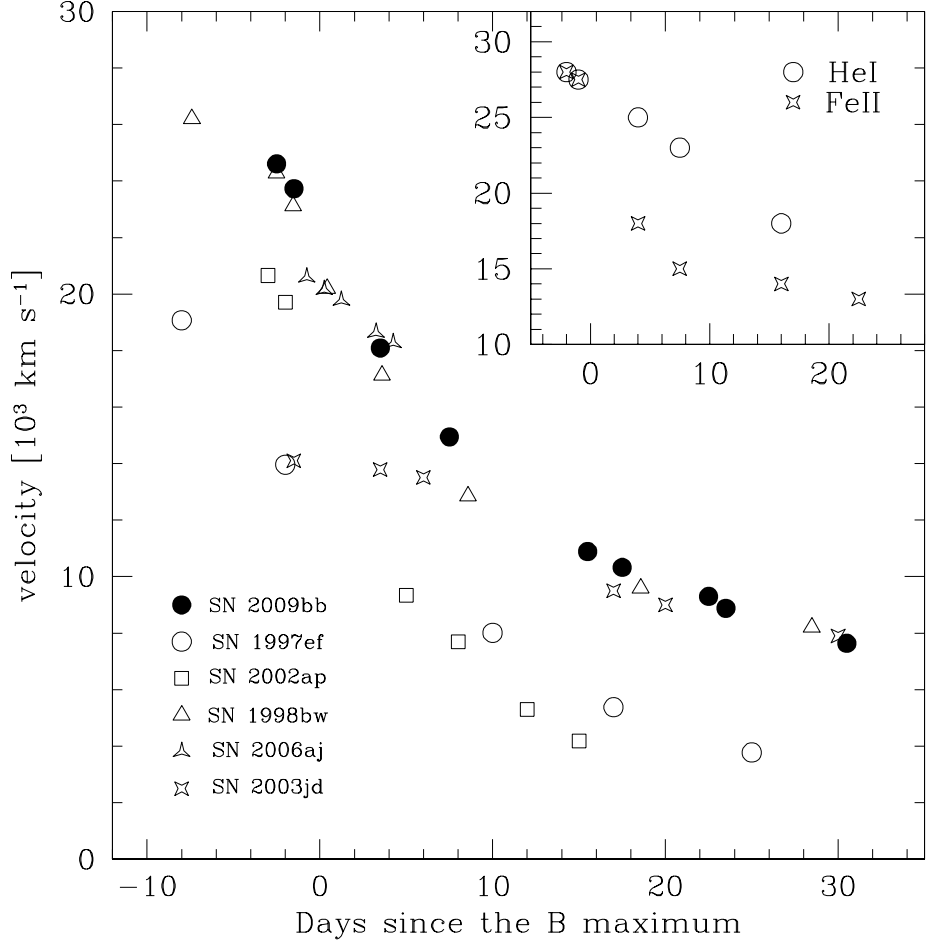}
\caption{Evolution of the velocity measured from the blueshift of the minimum of  \ion{Si}{2}~$\lambda$6355 line of SN~2009bb. For comparison the \ion{Si}{2}~$\lambda$6355 line velocities of SN~1997ef, SN~1998bw, SN~2002ap, SN~2003jd, and SN~2006aj are also reported. In the inserted plot, the velocities of the obtained fitting the \ion{Fe}{2} $\lambda\lambda$4924, 5018, 5169 lines with SYNOW are compared with those obtained fitting the \ion{He}{1} lines with the same code.}  
\label{fig4.8}
\end{figure}

\begin{figure}
\plotone{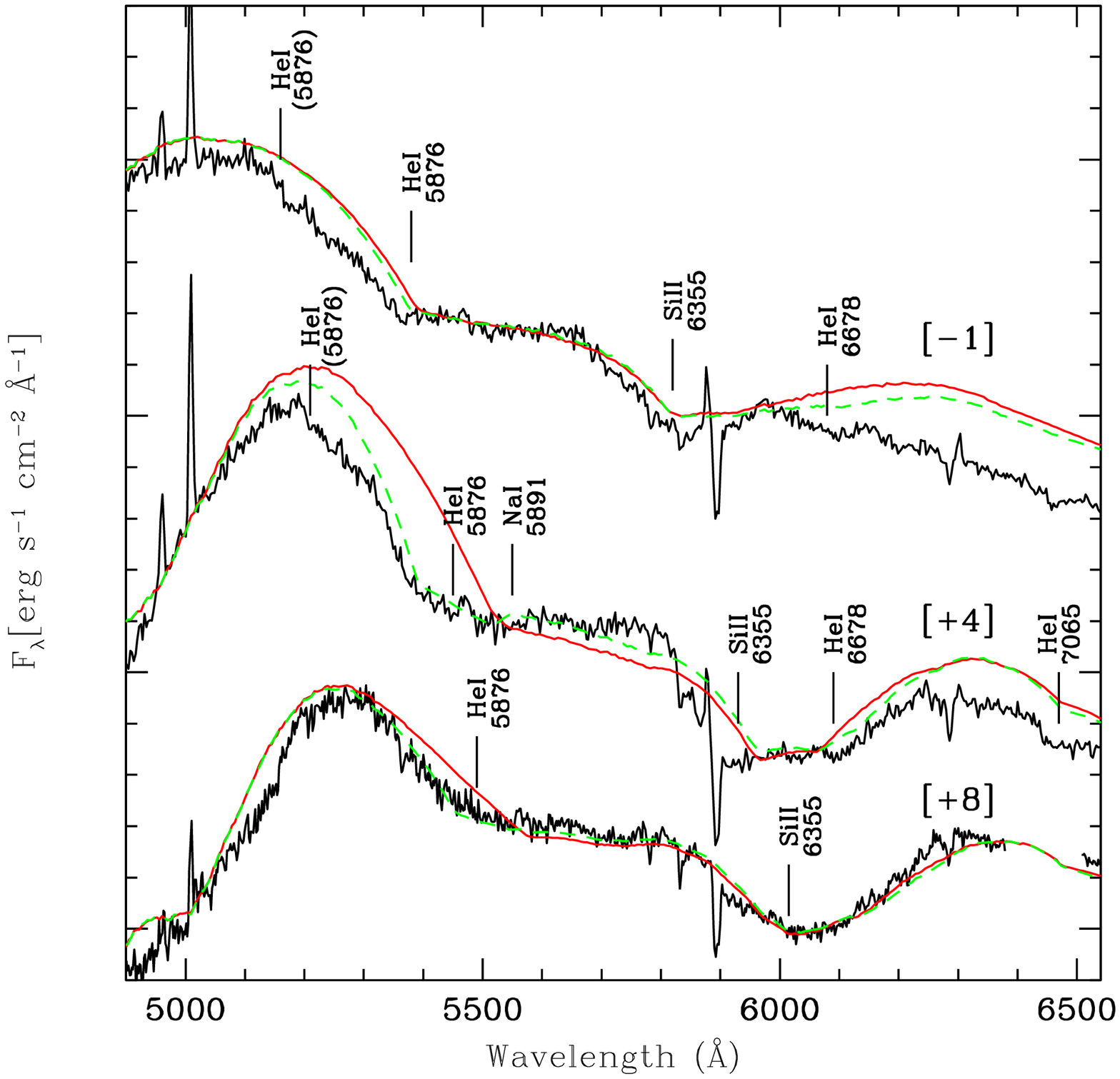}
\caption{Spectral models of  SN~2009bb obtained with SYNOW. The phase measured since the $B$ maximum brightness is reported in the plot. In the spectrum on top of the graph, the green  dashed line indicates the model with \ion{He}{1}, \ion{Fe}{2}, \ion{Si}{2}, while the red solid one have the same elements but with  \ion{Na}{1} instead of \ion{He}{1}. In the middle spectrum, the green  dashed line is a  model  with \ion{He}{1} (detached), \ion{Na}{1}, \ion{Fe}{2}, \ion{Si}{2} and  \ion{Ne}{1}, while the red solid line is the same model but without helium. Finally, in the bottom spectrum, the green  dashed line is a  model  with \ion{He}{1} (detached) \ion{Na}{1}, \ion{Fe}{2}, \ion{Si}{2} and  \ion{Ne}{1}, while the red solid line is the same model but without helium.}
\label{fig4.6}
\end{figure}

\begin{figure}
\includegraphics[scale=.60,angle=90]{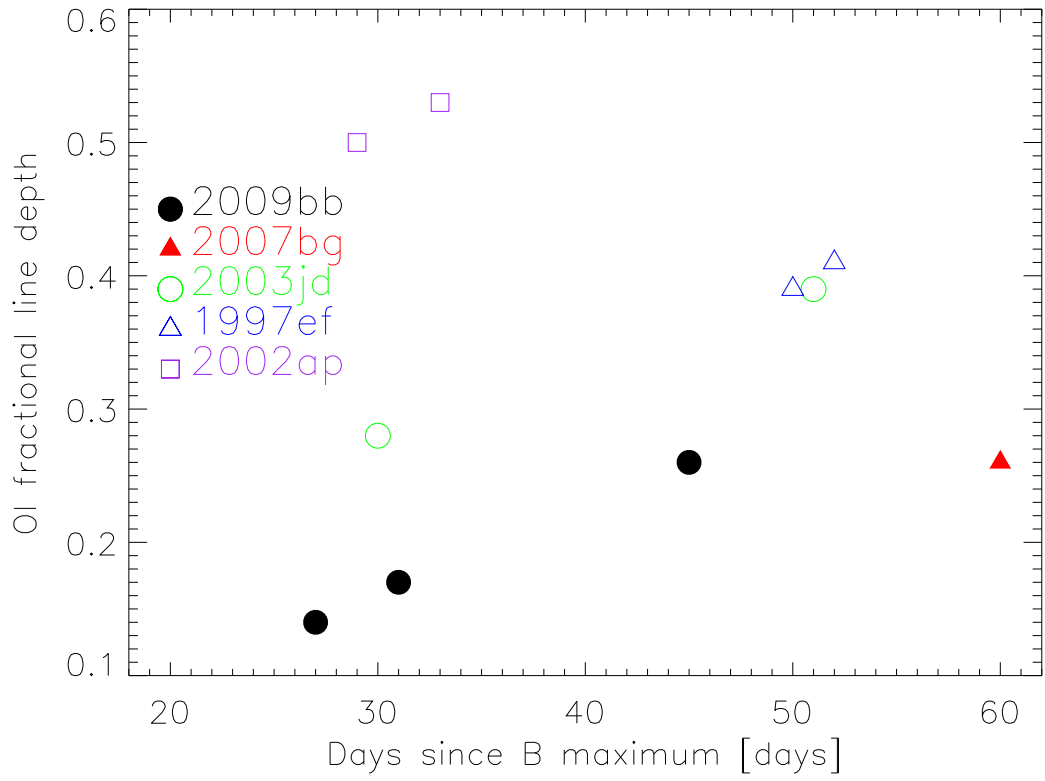}
\caption{Fractional line depth of the \ion{O}{1} $\lambda$7774 line measured using the formula of \citet{Matheson01}.}
\label{fig4.7}
\end{figure}

\begin{figure}
\plotone{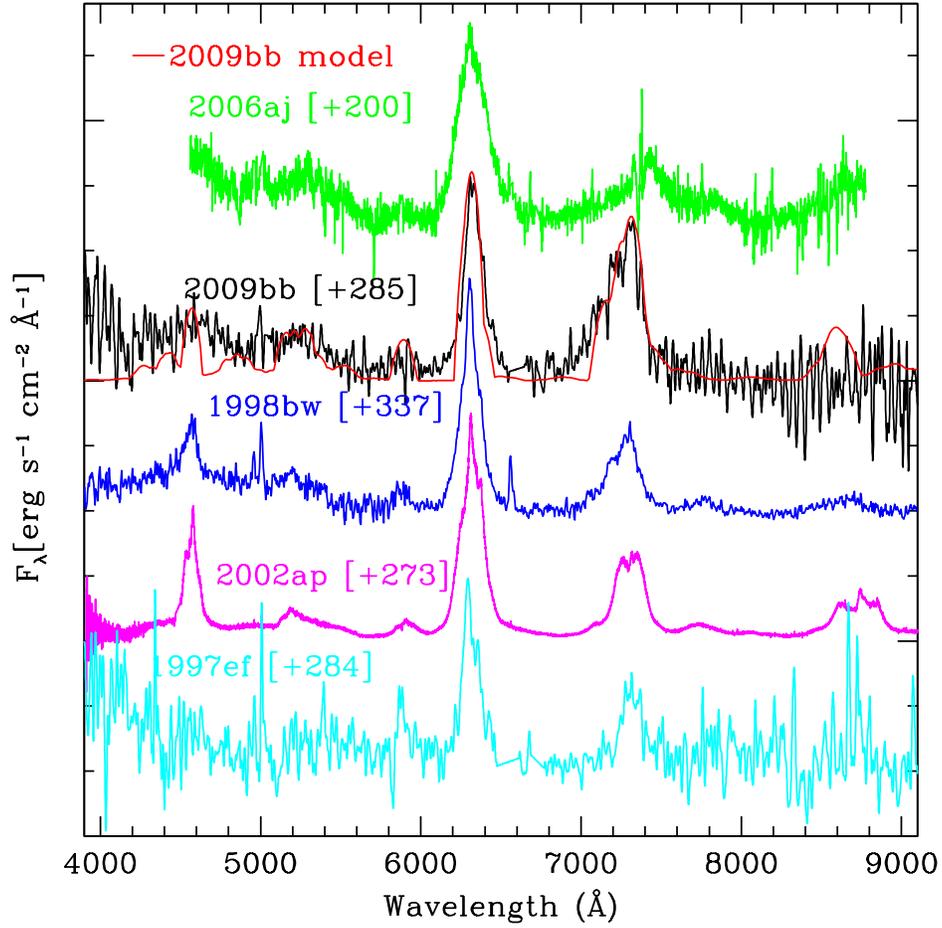}
\caption{Comparison between the  nebular spectra  of SN~2009bb, SN~1997ef, SN~1998bw, SN~2002ap and SN~2006aj. The model spectrum of  SN~2009bb discussed in Section 5.3 is also plotted.}
\label{fig4.9}
\end{figure}

\begin{figure}
\plotone{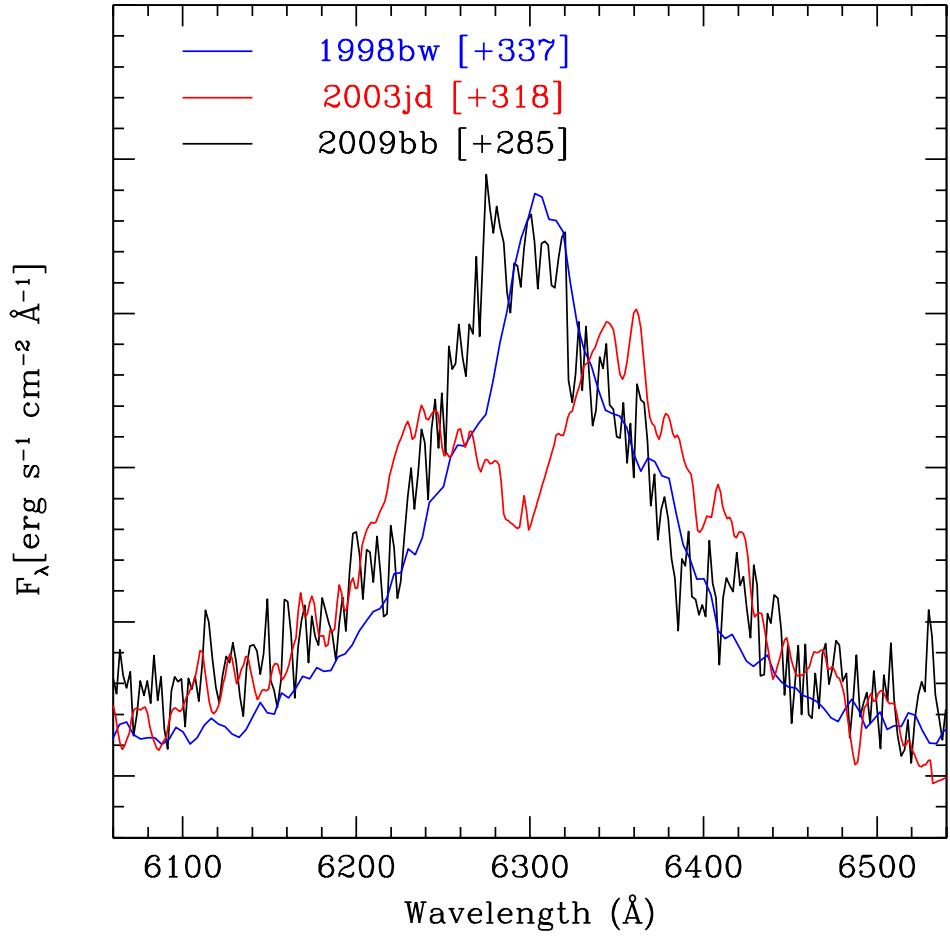}
\caption{Comparison between the [\ion{O}{1}] $\lambda\lambda$6300, 6354 nebular emission lines of SN~2009bb, SN~1998bw and SN~2003jd.}
\label{fig4.10}
\end{figure}








\clearpage

\begin{deluxetable}{cccccccc}
\tabletypesize{\scriptsize}
\tablecaption{$BVRI$ photometry of SN~2009bb}
\tablewidth{0pt}
\label{tab1}
\tablehead{
\colhead{date} & \colhead{J.D.} & \colhead{$B$} & \colhead{$V$} & \colhead{$R$} & \colhead{$I$} & \colhead{Telescope} 
}
\startdata
24/3/2009 & 2454914.5          & 18.10 $\pm$  0.06 &  $-$  &  $-$  &  $-$  & PROMPT3 \\
24/3/2009 & 2454914.5          &  $-$  & 17.21 $\pm$  0.06 & 17.09 $\pm$  0.03 & 16.75 $\pm$  0.08 & PROMPT5 \\
24/3/2009 & 2454914.6          & 18.00 $\pm$ 0.1 &  $-$  &  $-$  &  $-$  & Swope \\
25/3/2009 & 2454915.6          & 17.49 $\pm$  5.27 & 16.82 $\pm$ 0.1 &  $-$  &  $-$  & Swope \\
26/3/2009 & 2454916.5          & 17.33 $\pm$  0.04 &  $-$  &  $-$  &  $-$  & PROMPT3 \\
26/3/2009 & 2454916.5          &  $-$  & 16.65 $\pm$  0.04 & 16.36 $\pm$  0.03 & 15.95 $\pm$  0.05 & PROMPT5 \\
27/3/2009 & 2454917.5          & 17.21 $\pm$  0.03 &  $-$  &  $-$  &  $-$  & PROMPT3 \\
27/3/2009 & 2454917.5          &  $-$  & 16.48 $\pm$  0.04 & 16.15 $\pm$  0.03 & 15.80 $\pm$  0.05 & PROMPT5 \\
27/3/2009 & 2454917.7          & 17.17 $\pm$  0.04 & 16.40 $\pm$ 0.1 &  $-$  &  $-$  & Swope \\
28/3/2009 & 2454918.5          & 17.13 $\pm$  0.04 &  $-$  &  $-$  &  $-$  & PROMPT3 \\
28/3/2009 & 2454918.5          &  $-$  & 16.34 $\pm$  0.03 & 16.05 $\pm$  0.03 &  $-$  & PROMPT5 \\
28/3/2009 & 2454918.6          &  $-$  &  $-$  &  $-$  & 15.71 $\pm$  0.04 & PROMPT5 \\
29/3/2009 & 2454919.7          & 17.08 $\pm$  0.04 &  $-$  &  $-$  &  $-$  & PROMPT3 \\
29/3/2009 & 2454919.7          &  $-$  & 16.21 $\pm$  0.03 &  $-$  & 15.59 $\pm$  0.06 & PROMPT5 \\
30/3/2009 & 2454920.5          & 17.07 $\pm$  0.04 &  $-$  &  $-$  &  $-$  & PROMPT3 \\
30/3/2009 & 2454920.5          &  $-$  & 16.21 $\pm$  0.04 & 15.90 $\pm$  0.03 & 15.50 $\pm$  0.05 & PROMPT5 \\
30/3/2009 & 2454920.6          & 17.07 $\pm$  0.03 & 16.16 $\pm$ 0.1 &  $-$  &  $-$  & Swope \\
31/3/2009 & 2454921.5          &  $-$  & 16.10 $\pm$  0.03 & 15.85 $\pm$  0.03 & 15.49 $\pm$  0.05 & PROMPT5 \\
1/4/2009 & 2454922.5          & 17.07 $\pm$  0.03 &  $-$  &  $-$  &  $-$  & PROMPT3 \\
1/4/2009 & 2454922.5          &  $-$  & 16.08 $\pm$  0.03 & 15.81 $\pm$  0.03 & 15.42 $\pm$  0.06 & PROMPT5 \\
2/4/2009 & 2454923.6          &  $-$  &  $-$  &  $-$  & 15.35 $\pm$  0.04 & PROMPT5 \\
2/4/2009 & 2454923.7          & 17.16 $\pm$  0.04 &  $-$  &  $-$  &  $-$  & PROMPT3 \\
2/4/2009 & 2454923.7          &  $-$  & 16.07 $\pm$  0.03 & 15.79 $\pm$  0.03 &  $-$  & PROMPT5 \\
3/4/2009 & 2454924.5          & 17.25 $\pm$  0.04 &  $-$  &  $-$  &  $-$  & PROMPT3 \\
3/4/2009 & 2454924.6          &  $-$  & 16.08 $\pm$  0.03 & 15.78 $\pm$  0.03 & 15.35 $\pm$  0.06 & PROMPT5 \\
6/4/2009 & 2454927.6          & 17.57 $\pm$  0.27 & 16.21 $\pm$ 0.1 &  $-$  &  $-$  & Swope \\
7/4/2009 & 2454928.5          & 17.78 $\pm$  0.03 &  $-$  &  $-$  &  $-$  & PROMPT3 \\
7/4/2009 & 2454928.5          &  $-$  & 16.31 $\pm$  0.03 & 15.85 $\pm$  0.04 & 15.37 $\pm$  0.04 & PROMPT5 \\
7/4/2009 & 2454928.6          & 17.75 $\pm$  0.07 & 16.31 $\pm$ 0.1 &  $-$  &  $-$  & Swope \\
8/4/2009 & 2454929.5          & 17.84 $\pm$  0.04 &  $-$  &  $-$  &  $-$  & PROMPT3 \\
8/4/2009 & 2454929.5          &  $-$  & 16.41 $\pm$  0.04 & 15.90 $\pm$  0.03 & 15.37 $\pm$  0.05 & PROMPT5 \\
9/4/2009 & 2454930.5          & 18.00 $\pm$  0.07 & 16.50 $\pm$ 0.1 &  $-$  &  $-$  & Swope \\
9/4/2009 & 2454930.5          & 18.02 $\pm$  0.05 &  $-$  &  $-$  &  $-$  & PROMPT3 \\
9/4/2009 & 2454930.5          &  $-$  & 16.52 $\pm$  0.04 & 16.00 $\pm$  0.03 & 15.40 $\pm$  0.05 & PROMPT5 \\
10/4/2009 & 2454931.5          &  $-$  & 16.59 $\pm$  0.04 & 16.01 $\pm$  0.03 & 15.44 $\pm$  0.04 & PROMPT5 \\
10/4/2009 & 2454931.6          & 18.13 $\pm$  0.05 & 16.56 $\pm$ 0.1 &  $-$  &  $-$  & Swope \\
11/4/2009 & 2454932.5          &  $-$  & 16.71 $\pm$  0.04 & 16.10 $\pm$  0.03 & 15.48 $\pm$  0.05 & PROMPT5 \\
12/4/2009 & 2454933.5          &  $-$  & 16.76 $\pm$  0.05 & 16.14 $\pm$  0.03 &  $-$  & PROMPT5 \\
12/4/2009 & 2454933.6          & 18.40 $\pm$  0.10 & 16.75 $\pm$ 0.1 &  $-$  &  $-$  & Swope \\
13/4/2009 & 2454934.5          & 18.48 $\pm$  0.05 &  $-$  &  $-$  &  $-$  & PROMPT3 \\
13/4/2009 & 2454934.5          &  $-$  & 16.90 $\pm$  0.04 & 16.22 $\pm$  0.05 & 15.59 $\pm$  0.04 & PROMPT5 \\
14/4/2009 & 2454935.5          & 18.59 $\pm$  0.06 &  $-$  &  $-$  &  $-$  & PROMPT3 \\
14/4/2009 & 2454935.5          &  $-$  & 16.95 $\pm$  0.05 & 16.33 $\pm$  0.03 & 15.67 $\pm$  0.05 & PROMPT5 \\
15/4/2009 & 2454936.5          & 18.67 $\pm$  0.05 &  $-$  &  $-$  &  $-$  & PROMPT3 \\
15/4/2009 & 2454936.5          &  $-$  & 17.11 $\pm$  0.05 & 16.43 $\pm$  0.03 & 15.74 $\pm$  0.04 & PROMPT5 \\
16/4/2009 & 2454937.6          & 18.81 $\pm$  0.09 & 17.15 $\pm$ 0.1 &  $-$  &  $-$  & Swope \\
17/4/2009 & 2454938.5          &  $-$  & 17.25 $\pm$  0.06 & 16.60 $\pm$  0.03 & 15.89 $\pm$  0.05 & PROMPT5 \\
17/4/2009 & 2454938.6          & 18.92 $\pm$  0.14 & 17.22 $\pm$ 0.1 &  $-$  &  $-$  & Swope \\
18/4/2009 & 2454939.5          &  $-$  & 17.35 $\pm$  0.06 & 16.68 $\pm$  0.03 & 15.96 $\pm$  0.05 & PROMPT5 \\
20/4/2009 & 2454941.6          &  $-$  & 17.54 $\pm$  0.06 & 16.82 $\pm$  0.03 & 16.12 $\pm$  0.05 & PROMPT5 \\
21/4/2009 & 2454942.5          &  $-$  & 17.54 $\pm$  0.07 & 16.92 $\pm$  0.03 & 16.16 $\pm$  0.05 & PROMPT5 \\
21/4/2009 & 2454942.6          & 19.16 $\pm$  0.12 & 17.51 $\pm$ 0.1 &  $-$  &  $-$  & Swope \\
22/4/2009 & 2454943.5          &  $-$  & 17.62 $\pm$  0.07 & 16.96 $\pm$  0.04 & 16.23 $\pm$  0.05 & PROMPT5 \\
23/4/2009 & 2454944.6          &  $-$  & 17.67 $\pm$  0.06 &  $-$  & 16.38 $\pm$  0.05 & PROMPT5 \\
23/4/2009 & 2454944.7          & 19.38 $\pm$  0.10 &  $-$  &  $-$  &  $-$  & PROMPT3 \\
25/4/2009 & 2454946.5          & 19.38 $\pm$  0.09 &  $-$  &  $-$  &  $-$  & PROMPT3 \\
25/4/2009 & 2454946.5          &  $-$  & 17.74 $\pm$  0.07 &  $-$  &  $-$  & PROMPT5 \\
25/4/2009 & 2454946.6          &  $-$  &  $-$  & 17.18 $\pm$  0.03 & 16.45 $\pm$  0.06 & PROMPT5 \\
26/4/2009 & 2454947.5          &  $-$  & 17.80 $\pm$ 0.1 &  $-$  &  $-$  & Swope \\
26/4/2009 & 2454947.5          &  $-$  &  $-$  &  $-$  & 16.47 $\pm$  0.05 & PROMPT5 \\
26/4/2009 & 2454947.6          & 19.43 $\pm$  0.19 &  $-$  &  $-$  &  $-$  & Swope \\
26/4/2009 & 2454947.6          &  $-$  & 17.85 $\pm$  0.08 & 17.20 $\pm$  0.03 &  $-$  & PROMPT5 \\
27/4/2009 & 2454948.6          &  $-$  & 17.88 $\pm$  0.07 & 17.24 $\pm$  0.03 & 16.54 $\pm$  0.06 & PROMPT5 \\
29/4/2009 & 2454950.5          &  $-$  &  $-$  & 17.33 $\pm$  0.03 &  $-$  & PROMPT5 \\
29/4/2009 & 2454950.6          &  $-$  & 17.95 $\pm$  0.11 &  $-$  & 16.60 $\pm$  0.07 & PROMPT5 \\
30/4/2009 & 2454951.5          &  $-$  & 17.99 $\pm$  0.10 & 17.35 $\pm$  0.04 & 16.61 $\pm$  0.08 & PROMPT5 \\
1/5/2009 & 2454952.5          &  $-$  & 18.07 $\pm$  0.11 & 17.39 $\pm$  0.03 & 16.64 $\pm$  0.07 & PROMPT5 \\
2/5/2009 & 2454953.5          & 19.48 $\pm$  0.04 &  $-$  &  $-$  &  $-$  & PROMPT3 \\
3/5/2009 & 2454954.5          &  $-$  & 18.15 $\pm$  0.10 & 17.50 $\pm$  0.04 & 16.73 $\pm$  0.07 & PROMPT5 \\
4/5/2009 & 2454955.5          &  $-$  & 18.16 $\pm$  0.11 & 17.50 $\pm$  0.03 & 16.81 $\pm$  0.08 & PROMPT5 \\
5/5/2009 & 2454956.5          & 19.93 $\pm$  0.49 &  $-$  &  $-$  &  $-$  & PROMPT3 \\
5/5/2009 & 2454956.5          &  $-$  & 18.30 $\pm$  0.11 & 17.63 $\pm$  0.03 & 16.87 $\pm$  0.06 & PROMPT5 \\
8/5/2009 & 2454959.5          & 19.70 $\pm$  0.20 &  $-$  &  $-$  &  $-$  & PROMPT3 \\
8/5/2009 & 2454959.5          &  $-$  & 18.21 $\pm$  0.09 & 17.70 $\pm$  0.03 & 16.95 $\pm$  0.08 & PROMPT5 \\
9/5/2009 & 2454960.5          & 19.81 $\pm$  0.13 & 18.23 $\pm$ 0.1 &  $-$  &  $-$  & Swope \\
9/5/2009 & 2454960.5          & 19.88 $\pm$  0.14 &  $-$  &  $-$  &  $-$  & PROMPT3 \\
9/5/2009 & 2454960.5          &  $-$  & 18.21 $\pm$  0.11 & 17.72 $\pm$  0.03 &  $-$  & PROMPT5 \\
9/5/2009 & 2454960.6          &  $-$  &  $-$  &  $-$  & 16.86 $\pm$  0.08 & PROMPT5 \\
10/5/2009 & 2454961.6          & 20.08 $\pm$  0.16 &  $-$  &  $-$  &  $-$  & PROMPT3 \\
10/5/2009 & 2454961.6          &  $-$  & 18.36 $\pm$  0.12 & 17.79 $\pm$  0.03 & 17.06 $\pm$  0.10 & PROMPT5 \\
11/5/2009 & 2454962.5          & 19.86 $\pm$  0.27 &  $-$  &  $-$  &  $-$  & PROMPT3 \\
11/5/2009 & 2454962.5          &  $-$  & 18.31 $\pm$  0.14 & 17.71 $\pm$  0.04 & 16.94 $\pm$  0.07 & PROMPT5 \\
12/5/2009 & 2454963.5          & 19.64 $\pm$  0.27 &  $-$  &  $-$  &  $-$  & PROMPT3 \\
12/5/2009 & 2454963.5          &  $-$  & 18.41 $\pm$  0.14 & 17.95 $\pm$  0.03 & 17.08 $\pm$  0.05 & PROMPT5 \\
13/5/2009 & 2454964.5          & 19.91 $\pm$  0.29 &  $-$  &  $-$  &  $-$  & PROMPT3 \\
13/5/2009 & 2454964.5          &  $-$  & 18.75 $\pm$  0.17 & 17.90 $\pm$  0.03 &  $-$  & PROMPT5 \\
13/5/2009 & 2454964.6          & 19.96 $\pm$ 0.1 & 18.34 $\pm$ 0.1 &  $-$  &  $-$  & Swope \\
15/5/2009 & 2454966.5          & 20.03 $\pm$ 0.1 & 18.48 $\pm$ 0.1 &  $-$  &  $-$  & Swope \\
16/5/2009 & 2454967.5          &  $-$  & 18.29 $\pm$  0.09 & 17.99 $\pm$  0.03 & 16.93 $\pm$  0.09 & PROMPT5 \\
17/5/2009 & 2454968.5          &  $-$  & 18.46 $\pm$  0.13 & 17.98 $\pm$  0.04 & 17.21 $\pm$  0.11 & PROMPT5 \\
19/5/2009 & 2454970.5          &  $-$  & 18.49 $\pm$  0.15 & 17.98 $\pm$  0.06 & 17.39 $\pm$  0.12 & PROMPT5 \\
21/5/2009 & 2454972.5          &  $-$  & 18.99 $\pm$  0.18 & 18.14 $\pm$  0.03 & 17.33 $\pm$  0.11 & PROMPT5 \\
22/5/2009 & 2454973.5          &  $-$  & 18.72 $\pm$  0.18 & 18.11 $\pm$  0.03 & 17.35 $\pm$  0.11 & PROMPT5 \\
2/6/2009 & 2454984.5          &  $-$  & 18.95 $\pm$  0.19 & 18.57 $\pm$  0.05 &  $-$  & PROMPT5 \\
2/6/2009 & 2454984.6          &  $-$  &  $-$  &  $-$  & 17.47 $\pm$  0.17 & PROMPT5 \\
11/6/2009 & 2454993.5          &  $-$  & 19.14 $\pm$  0.17 & 18.58 $\pm$  0.03 & 18.11 $\pm$  0.20 & PROMPT5 \\
12/6/2009 & 2454994.5          &  $-$  &  $-$  & 18.90 $\pm$  0.03 &  $-$  & PROMPT5 \\
12/6/2009 & 2454994.6          &  $-$  &  $-$  &  $-$  & 18.00 $\pm$  0.23 & PROMPT5 \\
13/6/2009 & 2454995.5          &  $-$  & 19.59 $\pm$  0.29 & 18.84 $\pm$  0.04 & 18.00 $\pm$  0.17 & PROMPT5 \\
25/7/2009 & 2455037.5          &  $-$  & 19.78 $\pm$  0.78 &  $-$  &  $-$  & PROMPT5 \\
\enddata
\end{deluxetable}

\begin{deluxetable}{cccccccc}
\tabletypesize{\scriptsize}
\tablecaption{$u'g'r'i'z'$ photometry of SN~2009bb}
\tablewidth{0pt}
\label{tab1.2}
\tablehead{
\colhead{date} & \colhead{J.D.} & \colhead{$u'$} & \colhead{$g'$} & \colhead{$r'$} & \colhead{$i'$} & \colhead{$z'$} & \colhead{Telescope} 
}
\startdata
24/3/2009 & 2454914.5  &  $-$  &  $-$  & 17.33 $\pm$  0.03 & 17.24 $\pm$  0.05 & 17.12 $\pm$  0.08 & PROMPT5 \\
24/3/2009 & 2454914.5  &  $-$  & 17.58 $\pm$  0.06 &  $-$  &  $-$  &  $-$  & PROMPT3 \\
24/3/2009 & 2454914.6  & 18.84 $\pm$ 0.1 & 17.59 $\pm$ 0.1 & 17.23 $\pm$ 0.1 & 17.10 $\pm$ 0.1 &  $-$  & Swope \\
25/3/2009 & 2454915.6  & 18.43 $\pm$ 0.1 & 17.21 $\pm$  0.03 & 16.82 $\pm$  0.03 & 16.71 $\pm$  0.05 &  $-$  & Swope \\
26/3/2009 & 2454916.5  &  $-$  &  $-$  & 16.55 $\pm$  0.03 & 16.44 $\pm$  0.03 & 16.38 $\pm$  0.07 & PROMPT5 \\
26/3/2009 & 2454916.5  &  $-$  & 16.99 $\pm$  0.05 &  $-$  &  $-$  &  $-$  & PROMPT3 \\
27/3/2009 & 2454917.5  &  $-$  &  $-$  & 16.35 $\pm$  0.03 & 16.25 $\pm$  0.03 & 16.07 $\pm$  0.06 & PROMPT5 \\
27/3/2009 & 2454917.5  &  $-$  & 16.84 $\pm$  0.04 &  $-$  &  $-$  &  $-$  & PROMPT3 \\
27/3/2009 & 2454917.7  & 18.13 $\pm$ 0.1 & 16.82 $\pm$  0.03 & 16.32 $\pm$  0.03 & 16.27 $\pm$  0.04 &  $-$  & Swope \\
28/3/2009 & 2454918.5  &  $-$  &  $-$  & 16.20 $\pm$  0.03 & 16.18 $\pm$  0.03 & 15.95 $\pm$  0.07 & PROMPT5 \\
28/3/2009 & 2454918.5  &  $-$  & 16.72 $\pm$  0.04 &  $-$  &  $-$  &  $-$  & PROMPT3 \\
29/3/2009 & 2454919.7  &  $-$  & 16.66 $\pm$  0.04 &  $-$  &  $-$  &  $-$  & PROMPT3 \\
30/3/2009 & 2454920.5  &  $-$  &  $-$  & 16.06 $\pm$  0.03 & 16.03 $\pm$  0.03 & 15.80 $\pm$  0.06 & PROMPT5 \\
30/3/2009 & 2454920.5  &  $-$  & 16.62 $\pm$  0.04 &  $-$  &  $-$  &  $-$  & PROMPT3 \\
30/3/2009 & 2454920.6  & 18.11 $\pm$ 0.1 & 16.63 $\pm$  0.03 & 16.07 $\pm$  0.03 & 16.00 $\pm$  0.04 &  $-$  & Swope \\
31/3/2009 & 2454921.5  &  $-$  &  $-$  & 16.03 $\pm$  0.03 & 16.00 $\pm$  0.03 & 15.78 $\pm$  0.06 & PROMPT5 \\
1/4/2009 & 2454922.5  &  $-$  &  $-$  & 16.02 $\pm$  0.03 & 15.95 $\pm$  0.03 & 15.69 $\pm$  0.06 & PROMPT5 \\
1/4/2009 & 2454922.5  &  $-$  & 16.65 $\pm$  0.04 &  $-$  &  $-$  &  $-$  & PROMPT3 \\
2/4/2009 & 2454923.6  &  $-$  &  $-$  & 15.96 $\pm$  0.03 & 15.91 $\pm$  0.03 & 15.63 $\pm$  0.07 & PROMPT5 \\
2/4/2009 & 2454923.7  &  $-$  & 16.69 $\pm$  0.04 &  $-$  &  $-$  &  $-$  & PROMPT3 \\
3/4/2009 & 2454924.5  &  $-$  &  $-$  & 15.98 $\pm$  0.03 & 15.91 $\pm$  0.03 & 15.60 $\pm$  0.06 & PROMPT5 \\
3/4/2009 & 2454924.5  &  $-$  & 16.70 $\pm$  0.04 &  $-$  &  $-$  &  $-$  & PROMPT3 \\
6/4/2009 & 2454927.6  & 18.98 $\pm$ 0.1 & 16.97 $\pm$  0.03 & 16.02 $\pm$  0.04 & 15.90 $\pm$  0.03 &  $-$  & Swope \\
7/4/2009 & 2454928.5  & 19.27 $\pm$ 0.1 & 17.09 $\pm$  0.03 & 16.05 $\pm$  0.03 & 15.93 $\pm$  0.03 &  $-$  & Swope \\
7/4/2009 & 2454928.5  &  $-$  &  $-$  & 16.07 $\pm$  0.03 & 15.93 $\pm$  0.03 & 15.60 $\pm$  0.06 & PROMPT5 \\
7/4/2009 & 2454928.5  &  $-$  & 17.05 $\pm$  0.04 &  $-$  &  $-$  &  $-$  & PROMPT3 \\
8/4/2009 & 2454929.5  &  $-$  &  $-$  & 16.14 $\pm$  0.03 & 15.96 $\pm$  0.03 & 15.70 $\pm$  0.06 & PROMPT5 \\
8/4/2009 & 2454929.5  &  $-$  & 17.14 $\pm$  0.05 &  $-$  &  $-$  &  $-$  & PROMPT3 \\
9/4/2009 & 2454930.5  & 19.62 $\pm$ 0.1 & 17.30 $\pm$  0.03 & 16.16 $\pm$  0.04 & 16.00 $\pm$  0.04 &  $-$  & Swope \\
9/4/2009 & 2454930.5  &  $-$  &  $-$  & 16.15 $\pm$  0.03 & 16.00 $\pm$  0.04 & 15.64 $\pm$  0.07 & PROMPT5 \\
9/4/2009 & 2454930.5  &  $-$  & 17.30 $\pm$  0.05 &  $-$  &  $-$  &  $-$  & PROMPT3 \\
10/4/2009 & 2454931.5  & 19.71 $\pm$ 0.1 & 17.42 $\pm$  0.03 & 16.22 $\pm$  0.03 & 16.05 $\pm$  0.03 &  $-$  & Swope \\
10/4/2009 & 2454931.5  &  $-$  &  $-$  & 16.22 $\pm$  0.03 & 16.05 $\pm$  0.03 & 15.69 $\pm$  0.06 & PROMPT5 \\
10/4/2009 & 2454931.5  &  $-$  & 17.38 $\pm$  0.05 &  $-$  &  $-$  &  $-$  & PROMPT3 \\
11/4/2009 & 2454932.5  &  $-$  &  $-$  & 16.34 $\pm$  0.03 & 16.09 $\pm$  0.03 & 15.73 $\pm$  0.06 & PROMPT5 \\
11/4/2009 & 2454932.5  &  $-$  & 17.52 $\pm$  0.05 &  $-$  &  $-$  &  $-$  & PROMPT3 \\
12/4/2009 & 2454933.5  & 20.03 $\pm$ 0.1 & 17.67 $\pm$  0.03 & 16.37 $\pm$  0.03 & 16.15 $\pm$  0.04 &  $-$  & Swope \\
12/4/2009 & 2454933.5  &  $-$  &  $-$  & 16.40 $\pm$  0.03 &  $-$  & 15.77 $\pm$  0.06 & PROMPT5 \\
12/4/2009 & 2454933.5  &  $-$  & 17.62 $\pm$  0.05 &  $-$  &  $-$  &  $-$  & PROMPT3 \\
13/4/2009 & 2454934.5  &  $-$  &  $-$  & 16.48 $\pm$  0.03 & 16.25 $\pm$  0.03 & 15.85 $\pm$  0.06 & PROMPT5 \\
13/4/2009 & 2454934.5  &  $-$  & 17.72 $\pm$  0.07 &  $-$  &  $-$  &  $-$  & PROMPT3 \\
14/4/2009 & 2454935.5  &  $-$  &  $-$  & 16.55 $\pm$  0.03 & 16.30 $\pm$  0.04 & 15.90 $\pm$  0.06 & PROMPT5 \\
14/4/2009 & 2454935.5  &  $-$  & 17.92 $\pm$  0.07 &  $-$  &  $-$  &  $-$  & PROMPT3 \\
15/4/2009 & 2454936.5  &  $-$  &  $-$  & 16.69 $\pm$  0.03 & 16.41 $\pm$  0.04 & 15.96 $\pm$  0.06 & PROMPT5 \\
15/4/2009 & 2454936.5  &  $-$  & 18.07 $\pm$  0.08 &  $-$  &  $-$  &  $-$  & PROMPT3 \\
16/4/2009 & 2454937.6  & 20.51 $\pm$ 0.1 & 18.15 $\pm$  0.03 & 16.75 $\pm$  0.03 & 16.49 $\pm$  0.04 &  $-$  & Swope \\
17/4/2009 & 2454938.5  &  $-$  &  $-$  & 16.84 $\pm$  0.03 & 16.61 $\pm$  0.04 & 16.10 $\pm$  0.06 & PROMPT5 \\
17/4/2009 & 2454938.6  & 20.48 $\pm$ 0.1 & 18.25 $\pm$  0.03 & 16.86 $\pm$  0.03 & 16.58 $\pm$  0.03 &  $-$  & Swope \\
18/4/2009 & 2454939.5  &  $-$  &  $-$  & 16.96 $\pm$  0.03 & 16.63 $\pm$  0.04 & 16.16 $\pm$  0.06 & PROMPT5 \\
18/4/2009 & 2454939.5  &  $-$  & 18.30 $\pm$  0.10 &  $-$  &  $-$  &  $-$  & PROMPT3 \\
20/4/2009 & 2454941.6  &  $-$  &  $-$  & 17.10 $\pm$  0.03 & 16.81 $\pm$  0.04 & 16.31 $\pm$  0.07 & PROMPT5 \\
20/4/2009 & 2454941.6  &  $-$  & 18.55 $\pm$  0.13 &  $-$  &  $-$  &  $-$  & PROMPT3 \\
21/4/2009 & 2454942.5  &  $-$  &  $-$  & 17.23 $\pm$  0.03 & 16.83 $\pm$  0.04 & 16.36 $\pm$  0.07 & PROMPT5 \\
21/4/2009 & 2454942.5  &  $-$  & 18.62 $\pm$  0.13 &  $-$  &  $-$  &  $-$  & PROMPT3 \\
21/4/2009 & 2454942.6  & 20.65 $\pm$ 0.1 & 18.55 $\pm$  0.03 & 17.16 $\pm$  0.03 & 16.89 $\pm$  0.06 &  $-$  & Swope \\
22/4/2009 & 2454943.5  &  $-$  &  $-$  & 17.27 $\pm$  0.03 & 16.94 $\pm$  0.04 & 16.41 $\pm$  0.07 & PROMPT5 \\
22/4/2009 & 2454943.5  &  $-$  & 18.67 $\pm$  0.13 &  $-$  &  $-$  &  $-$  & PROMPT3 \\
23/4/2009 & 2454944.6  &  $-$  &  $-$  & 17.38 $\pm$  0.04 &  $-$  & 16.52 $\pm$  0.07 & PROMPT5 \\
23/4/2009 & 2454944.6  &  $-$  & 18.80 $\pm$  0.15 &  $-$  &  $-$  &  $-$  & PROMPT3 \\
25/4/2009 & 2454946.5  &  $-$  & 18.83 $\pm$  0.16 &  $-$  &  $-$  &  $-$  & PROMPT3 \\
25/4/2009 & 2454946.6  &  $-$  &  $-$  & 17.47 $\pm$  0.03 & 17.08 $\pm$  0.04 & 16.53 $\pm$  0.07 & PROMPT5 \\
26/4/2009 & 2454947.5  & 21.05 $\pm$ 0.1 & 18.98 $\pm$  0.06 & 17.54 $\pm$  0.03 & 17.24 $\pm$  0.05 &  $-$  & Swope \\
26/4/2009 & 2454947.5  &  $-$  &  $-$  & 17.48 $\pm$  0.03 & 17.21 $\pm$  0.05 & 16.58 $\pm$  0.07 & PROMPT5 \\
26/4/2009 & 2454947.5  &  $-$  & 18.98 $\pm$  0.14 &  $-$  &  $-$  &  $-$  & PROMPT3 \\
27/4/2009 & 2454948.5  &  $-$  & 18.98 $\pm$  0.18 &  $-$  &  $-$  &  $-$  & PROMPT3 \\
27/4/2009 & 2454948.6  &  $-$  &  $-$  & 17.56 $\pm$  0.04 & 17.24 $\pm$  0.05 & 16.68 $\pm$  0.07 & PROMPT5 \\
28/4/2009 & 2454949.5  &  $-$  & 19.06 $\pm$  0.19 &  $-$  &  $-$  &  $-$  & PROMPT3 \\
29/4/2009 & 2454950.5  &  $-$  &  $-$  &  $-$  &  $-$  & 16.71 $\pm$  0.08 & PROMPT5 \\
29/4/2009 & 2454950.6  &  $-$  &  $-$  & 17.69 $\pm$  0.04 & 17.38 $\pm$  0.05 &  $-$  & PROMPT5 \\
29/4/2009 & 2454950.6  &  $-$  & 19.12 $\pm$  0.20 &  $-$  &  $-$  &  $-$  & PROMPT3 \\
30/4/2009 & 2454951.5  &  $-$  &  $-$  & 17.75 $\pm$  0.04 & 17.43 $\pm$  0.05 & 16.79 $\pm$  0.08 & PROMPT5 \\
30/4/2009 & 2454951.5  &  $-$  & 19.29 $\pm$  0.20 &  $-$  &  $-$  &  $-$  & PROMPT3 \\
1/5/2009 & 2454952.5  &  $-$  &  $-$  & 17.83 $\pm$  0.04 & 17.50 $\pm$  0.05 & 16.79 $\pm$  0.08 & PROMPT5 \\
1/5/2009 & 2454952.5  &  $-$  & 19.32 $\pm$  0.23 &  $-$  &  $-$  &  $-$  & PROMPT3 \\
2/5/2009 & 2454953.5  &  $-$  & 19.26 $\pm$  0.18 &  $-$  &  $-$  &  $-$  & PROMPT3 \\
3/5/2009 & 2454954.5  &  $-$  &  $-$  & 17.88 $\pm$  0.03 & 17.61 $\pm$  0.06 & 16.90 $\pm$  0.08 & PROMPT5 \\
3/5/2009 & 2454954.5  &  $-$  & 19.37 $\pm$  0.17 &  $-$  &  $-$  &  $-$  & PROMPT3 \\
4/5/2009 & 2454955.5  &  $-$  &  $-$  & 17.93 $\pm$  0.04 & 17.57 $\pm$  0.05 & 16.85 $\pm$  0.09 & PROMPT5 \\
5/5/2009 & 2454956.5  &  $-$  &  $-$  & 17.97 $\pm$  0.06 &  $-$  & 16.96 $\pm$  1.84 & PROMPT5 \\
8/5/2009 & 2454959.5  &  $-$  &  $-$  & 18.17 $\pm$  0.05 & 17.70 $\pm$  0.06 & 17.08 $\pm$  0.10 & PROMPT5 \\
8/5/2009 & 2454959.5  &  $-$  & 19.46 $\pm$  0.20 &  $-$  &  $-$  &  $-$  & PROMPT3 \\
9/5/2009 & 2454960.5  & 21.62 $\pm$ 0.1 & 19.49 $\pm$  0.03 & 18.20 $\pm$  0.07 & 17.80 $\pm$  0.07 &  $-$  & Swope \\
9/5/2009 & 2454960.5  &  $-$  &  $-$  & 18.18 $\pm$  0.06 & 17.69 $\pm$  0.05 &  $-$  & PROMPT5 \\
9/5/2009 & 2454960.5  &  $-$  & 19.43 $\pm$  0.46 &  $-$  &  $-$  &  $-$  & PROMPT3 \\
9/5/2009 & 2454960.6  &  $-$  &  $-$  &  $-$  &  $-$  & 17.04 $\pm$  0.08 & PROMPT5 \\
10/5/2009 & 2454961.5  &  $-$  &  $-$  & 18.12 $\pm$  0.04 &  $-$  &  $-$  & PROMPT5 \\
10/5/2009 & 2454961.6  &  $-$  &  $-$  &  $-$  & 17.84 $\pm$  0.07 & 17.11 $\pm$  0.07 & PROMPT5 \\
10/5/2009 & 2454961.6  &  $-$  & 19.53 $\pm$  0.28 &  $-$  &  $-$  &  $-$  & PROMPT3 \\
11/5/2009 & 2454962.5  &  $-$  &  $-$  &  $-$  & 17.86 $\pm$  0.06 & 17.07 $\pm$  0.09 & PROMPT5 \\
12/5/2009 & 2454963.5  &  $-$  &  $-$  & 18.19 $\pm$  0.04 & 17.89 $\pm$  0.05 & 17.17 $\pm$  0.11 & PROMPT5 \\
13/5/2009 & 2454964.5  &  $-$  & 19.51 $\pm$  0.33 &  $-$  &  $-$  &  $-$  & PROMPT3 \\
13/5/2009 & 2454964.6  & 21.61 $\pm$ 0.1 & 19.59 $\pm$  0.04 & 18.30 $\pm$  0.05 & 17.96 $\pm$  0.10 &  $-$  & Swope \\
15/5/2009 & 2454966.5  &  $-$  & 19.65 $\pm$  0.15 & 18.43 $\pm$  0.19 & 18.00 $\pm$  0.07 &  $-$  & Swope \\
17/5/2009 & 2454968.5  &  $-$  & 19.88 $\pm$  0.70 &  $-$  &  $-$  &  $-$  & PROMPT3 \\
21/5/2009 & 2454972.5  &  $-$  &  $-$  & 18.60 $\pm$  0.06 & 18.22 $\pm$  0.08 & 17.48 $\pm$  0.10 & PROMPT5 \\
22/5/2009 & 2454973.6  &  $-$  &  $-$  & 18.51 $\pm$  0.05 & 18.29 $\pm$  0.09 & 17.53 $\pm$  0.10 & PROMPT5 \\
22/5/2009 & 2454973.6  &  $-$  & 19.95 $\pm$  0.48 &  $-$  &  $-$  &  $-$  & PROMPT3 \\
23/5/2009 & 2454974.5  &  $-$  &  $-$  & 18.69 $\pm$  0.07 & 18.25 $\pm$  0.07 & 17.63 $\pm$  0.13 & PROMPT5 \\
\enddata
\end{deluxetable}

\begin{table}
\begin{center}
\caption{$YJH$ photometry of SN~2009bb}
\label{tab1.3}
\begin{tabular}{@{}cccccc}
\hline date & J.D. & $Y$ & $J$ & $H$ & Instr. \\
\hline
31/3/2009 & 2454921.7  & 15.21 $\pm$  0.04 & 15.08 $\pm$  0.04 & 14.89 $\pm$  0.04 &   RetroCam \\
8/4/2009 & 2454929.7  & 15.02 $\pm$  0.04 & 14.85 $\pm$  0.04 & 14.70 $\pm$  0.04 &    RetroCam \\
11/4/2009 & 2454932.7  & 15.06 $\pm$  0.04 & 14.92 $\pm$  0.04 & 14.80 $\pm$  0.04 &   RetroCam \\
14/4/2009 & 2454935.7  & 15.10 $\pm$  0.04 & 15.11 $\pm$  0.04 & 14.80 $\pm$  0.04 &   RetroCam \\
16/4/2009 & 2454937.7  &  $-$  &  $-$  & 14.94 $\pm$  0.04 &   RetroCam \\
18/4/2009 & 2454939.7  & 15.38 $\pm$  0.04 & 15.23 $\pm$  0.04 & 15.11 $\pm$  0.04 &   RetroCam \\
24/4/2009 & 2454945.7  & 15.81 $\pm$  0.04 & 15.65 $\pm$  0.04 & 15.47 $\pm$  0.04 &   RetroCam \\
27/4/2009 & 2454948.7  & 15.96 $\pm$  0.04 & 15.84 $\pm$  0.04 & 15.68 $\pm$  0.05 &   RetroCam \\
28/4/2009 & 2454949.7  & 16.01 $\pm$  0.04 & 15.95 $\pm$  0.04 & 15.66 $\pm$  0.05 &   RetroCam \\
24/5/2009 & 2454975.7  & 17.05 $\pm$  0.04 & 17.23 $\pm$  0.05 & 16.68 $\pm$  0.08 &   RetroCam \\

\hline
\end{tabular}
\end{center}
\end{table}

\begin{table}
\caption{Optical peak light curve parameters }
\label{tab1.4}
\hspace{15pt}\\
\begin{tabular}{@{}cccc}

 \hline
  Filter & JD max$^a$ & m(obs)$^b$ &  M(abs)$^c$ \\
  \hline
$B$  &  2454921.0 $\pm$ 0.5 & 17.05 $\pm$ 0.02 & $-$18.36 $\pm$ 0.44 \\
$V$  &  2454923.5 $\pm$ 0.5 & 16.16 $\pm$ 0.02 & $-$18.65 $\pm$ 0.34 \\
$R$  &  2454924.1 $\pm$ 0.5 & 15.80 $\pm$ 0.01 & $-$18.56 $\pm$ 0.28 \\
$I$  &  2454926.5 $\pm$ 0.5 & 15.36 $\pm$ 0.03 & $-$18.51 $\pm$ 0.21 \\
$u'$ &  2454919.6 $\pm$ 0.5 & 18.04 $\pm$ 0.08 & $-$17.68 $\pm$ 0.50 \\
$g'$ &  2454921.9 $\pm$ 0.5 & 16.62 $\pm$ 0.02 & $-$18.53 $\pm$ 0.40 \\
$r'$ &  2454924.4 $\pm$ 0.5 & 15.96 $\pm$ 0.02 & $-$18.60 $\pm$ 0.31 \\
$i'$ &  2454925.5 $\pm$ 0.5 & 15.89 $\pm$ 0.02 & $-$18.28 $\pm$ 0.25 \\ 
$z'$ &  2454926.7 $\pm$ 0.5 & 15.60 $\pm$ 0.02 & $-$18.27 $\pm$ 0.21 \\
$Y$  &  2454929.8 $\pm$ 1.0 & 15.01 $\pm$ 0.05 & $-$18.58 $\pm$ 0.19 \\
$J $ &  2454928.4 $\pm$ 1.0 & 14.85 $\pm$ 0.05 & $-$18.67 $\pm$ 0.18 \\
$H $ &  2454929.2 $\pm$ 2.0 & 14.71 $\pm$ 0.06 & $-$18.64 $\pm$ 0.17 \\

\hline
\end{tabular}
\\
$^a$ JD time of peak brightness.\\
$^b$ Apparent magnitude.\\
$^c$ Absolute magnitude.\\  
\end{table}

\begin{table}

\caption{Journal of spectroscopic observations}
\begin{tabular}{ccccccc}

\hline
UT date & J.D. & \multicolumn{1}{c}{epoch$^a$} &
\multicolumn{1}{c}{range}  & \multicolumn{1}{c}{res.} &  exptime & Instrument \\
     &        & \multicolumn{1}{c}{(days)}    &
     \multicolumn{1}{c}{(\AA)}  & \multicolumn{1}{c}{(\AA)}&  \multicolumn{1}{c}{(sec)}&       \\
\hline
  28/03/09 & 2454918.7   & $-$2.3 &   3800$-$9200   & 3.0  & 700    & WFCCD \\
  29/03/09 & 2454919.6   & $-$1.4 &   3800$-$9200   & 3.0  & 900    & WFCCD\\
  03/04/09 & 2454924.6   &  3.6   &   3800$-$9200   & 3.0  & 700    & WFCCD \\ 
  07/04/09 & 2454928.5   &  7.5   &   4200$-$10000  & 1.9  & 400    & IMACS \\
  15/04/09 & 2454936.6   &  15.6  &   4200$-$8100   & 1.35 & 900    & GMOS \\
  17/04/09 & 2454938.6   &  17.6  &   3700$-$9300   & 2.0  & 600    & LDSS3 \\
  18/04/09 & 2454939.6   &  18.6  &   3400$-$9500   & 3.0  & 700    & B\&C \\ 
  22/04/09 & 2454943.6   &  22.6  &   3400$-$9500   & 3.0  & 900    & B\&C\\
  23/04/09 & 2454944.6   &  23.6  &   3400$-$9500   & 3.0  & 1000   & B\&C \\
  26/04/09 & 2454947.6   &  26.6  &   3900$-$8100   & 1.35 & 900    & GMOS \\
  30/04/09 & 2454938.6   &  30.6  &   3700$-$9300   & 2.0  & 600    & LDSS3 \\
  14/05/09 & 2454965.5   &  44.5  &   4400$-$10000  & 1.9  & 600    & IMACS \\
  09/01/10 & 2455205.7   &  285.0 &   3700$-$9300   & 2.0  & 3x1800 & LDSS3 \\
\hline
\end{tabular}\\
$^a$Relative to $B$-band maximum.  \\
\label{tab5.1}
\end{table}

\begin{table}
\caption{Mass of the elements used to generate the synthetic spectra in Figure~\ref{fig4.9}.}
\label{tab5.2}
\hspace{15pt}\\
\begin{tabular}{@{}cc}

 \hline
  Element & mass (M$_{\sun}$)\\
  \hline
O  &  1.06 \\
Mg & 0.012 \\
Si & 0.07  \\
S  & 0.23  \\
Ca & 0.05 \\ 
Ni & 0.25  \\

\hline
\end{tabular}
\\
\end{table}

\begin{deluxetable}{cccccccc}
\tabletypesize{\scriptsize}
\tablecaption{$BVRI$ photometry of the local photometric sequence}
\tablewidth{0pt}
\label{tab7.1}
\tablehead{
\colhead{id} & \colhead{R.A.} & \colhead{Dec.} & \colhead{$B$} & \colhead{$V$} & \colhead{$R$} & \colhead{$I$}
}
\startdata
 1 & 10:31:45.355 & -40:00:42.73 & 15.26 $\pm$  0.01 & 14.54 $\pm$  0.04 & 14.13 $\pm$  0.02 & 13.76 $\pm$  0.01 \\
 2 & 10:31:44.107 & -39:55:25.68 & 15.44 $\pm$  0.02 & 14.47 $\pm$  0.04 & 13.93 $\pm$  0.03 & 13.41 $\pm$  0.02 \\
 3 & 10:31:42.905 & -39:57:51.19 & 13.83 $\pm$  0.03 & 13.18 $\pm$  0.03 & 12.80 $\pm$  0.02 & 12.45 $\pm$  0.01 \\
 4 & 10:31:41.494 & -39:57:10.26 & 15.50 $\pm$  0.02 & 14.53 $\pm$  0.04 & 13.99 $\pm$  0.02 & 13.49 $\pm$  0.02 \\
 5 & 10:31:36.185 & -39:54:45.04 & 14.98 $\pm$  0.01 & 14.36 $\pm$  0.03 & 13.99 $\pm$  0.02 & 13.63 $\pm$  0.02 \\
 6 & 10:31:24.828 & -39:58:30.29 & 14.63 $\pm$  0.01 & 14.12 $\pm$  0.03 & 13.77 $\pm$  0.01 & 13.42 $\pm$  0.01 \\
 7 & 10:31:24.547 & -40:01:44.72 & 15.00 $\pm$  0.01 & 14.48 $\pm$  0.04 & 14.14 $\pm$  0.02 & 13.79 $\pm$  0.01 \\
 8 & 10:31:50.942 & -39:57:50.11 & 15.90 $\pm$  0.01 & 15.04 $\pm$  0.01 & 14.51 $\pm$  0.03 & 14.02 $\pm$  0.01 \\
 9 & 10:31:20.933 & -39:59:52.55 & 15.83 $\pm$  0.02 & 15.02 $\pm$  0.01 & 14.57 $\pm$  0.03 & 14.15 $\pm$  0.01 \\
 10 & 10:31:48.010 & -39:54:23.65 & 15.52 $\pm$  0.01 & 14.96 $\pm$  0.01 & 14.64 $\pm$  0.02 & 14.29 $\pm$  0.02 \\
 11 & 10:31:33.866 & -40:00:09.68 & 15.86 $\pm$  0.02 & 15.09 $\pm$  0.01 & 14.69 $\pm$  0.02 & 14.29 $\pm$  0.01 \\
 12 & 10:31:26.273 & -39:56:48.23 & 15.75 $\pm$  0.01 & 15.10 $\pm$  0.01 & 14.75 $\pm$  0.01 & 14.38 $\pm$  0.02 \\
 13 & 10:31:45.754 & -39:56:32.17 & 15.84 $\pm$  0.01 & 15.21 $\pm$  0.01 & 14.86 $\pm$  0.02 & 14.51 $\pm$  0.03 \\
 14 & 10:31:51.468 & -39:54:42.59 & 16.25 $\pm$  0.01 & 15.52 $\pm$  0.01 & 15.10 $\pm$  0.03 & 14.68 $\pm$  0.03 \\
 15 & 10:31:51.929 & -39:54:27.94 & 16.36 $\pm$  0.01 & 15.56 $\pm$  0.01 & 15.11 $\pm$  0.02 & 14.63 $\pm$  0.04 \\
 16 & 10:31:24.012 & -39:59:29.94 & 16.41 $\pm$  0.02 & 15.62 $\pm$  0.01 & 15.20 $\pm$  0.03 & 14.78 $\pm$  0.02 \\
 17 & 10:31:14.064 & -39:58:45.55 & 16.45 $\pm$  0.03 & 15.64 $\pm$  0.01 & 15.19 $\pm$  0.03 & 14.77 $\pm$  0.01 \\
 18 & 10:31:16.582 & -39:56:39.70 & 16.13 $\pm$  0.04 & 15.62 $\pm$  0.04 & 15.37 $\pm$  0.05 & 15.06 $\pm$  0.06 \\
 19 & 10:31:18.864 & -39:57:51.30 & 16.55 $\pm$  0.01 & 15.76 $\pm$  0.01 & 15.33 $\pm$  0.02 & 14.91 $\pm$  0.01 \\
 20 & 10:31:23.846 & -39:54:50.40 & 16.49 $\pm$  0.02 & 15.80 $\pm$  0.01 & 15.40 $\pm$  0.02 & 15.01 $\pm$  0.04 \\
 21 & 10:31:24.442 & -39:55:08.33 & 16.79 $\pm$  0.03 & 15.96 $\pm$  0.01 & 15.49 $\pm$  0.01 & 15.07 $\pm$  0.03 \\
 22 & 10:31:46.673 & -39:58:40.19 & 16.98 $\pm$  0.01 & 16.28 $\pm$  0.01 & 15.89 $\pm$  0.02 & 15.50 $\pm$  0.02 \\
 23 & 10:31:31.198 & -39:53:32.46 & 17.52 $\pm$  0.02 & 16.62 $\pm$  0.01 & 16.06 $\pm$  0.05 & 15.56 $\pm$  0.02 \\
 24 & 10:31:49.822 & -40:00:16.78 & 18.60 $\pm$  0.01 & 17.08 $\pm$  0.03 & 16.12 $\pm$  0.06 & 15.04 $\pm$  0.02 \\
 25 & 10:31:42.703 & -39:58:20.78 & 17.56 $\pm$  0.01 & 16.90 $\pm$  0.01 & 16.50 $\pm$  0.03 & 16.06 $\pm$  0.03 \\
 26 & 10:31:48.434 & -40:01:29.93 & 17.84 $\pm$  0.02 & 17.14 $\pm$  0.01 & 16.72 $\pm$  0.05 & 16.32 $\pm$  0.05 \\
 27 & 10:31:54.881 & -39:56:16.58 & 19.25 $\pm$  0.02 & 17.78 $\pm$  0.01 & 16.77 $\pm$  0.14 & 15.98 $\pm$  0.10 \\
 28 & 10:31:33.463 & -39:58:54.05 & 18.11 $\pm$  0.02 & 17.45 $\pm$  0.01 & 17.10 $\pm$  0.04 & 16.71 $\pm$  0.02 \\
 29 & 10:31:37.385 & -39:57:52.78 & 18.83 $\pm$  0.02 & 17.86 $\pm$  0.01 & 17.35 $\pm$  0.09 & 16.82 $\pm$  0.04 \\
 30 & 10:31:32.995 & -39:58:26.69 & 19.39 $\pm$  0.02 & 18.22 $\pm$  0.01 & 17.52 $\pm$  0.04 & 16.76 $\pm$  0.21 \\
 31 & 10:31:31.990 & -39:58:28.85 & 18.86 $\pm$  0.01 & 18.11 $\pm$  0.01 & 17.64 $\pm$  0.06 & 17.21 $\pm$  0.06 \\
\enddata
\end{deluxetable}

\begin{deluxetable}{ccccccccc}
\tabletypesize{\scriptsize}
\tablecaption{$u'g'r'i'z'$ photometry of the local photometric sequence}
\tablewidth{0pt}
\label{tab7.2}
\tablehead{
\colhead{id} & \colhead{R.A.} & \colhead{Dec.} & \colhead{$u'$} & \colhead{$g'$} & \colhead{$r'$} & \colhead{$i'$} & \colhead{$z'$}
}
\startdata
 1 & 10:31:45.355 & -40:00:42.73 & 16.37 $\pm$  0.06 & 14.83 $\pm$  0.01 & 14.31 $\pm$  0.02 & 14.14 $\pm$  0.03 & 14.09 $\pm$  0.02 \\
 2 & 10:31:44.107 & -39:55:25.68 & 17.04 $\pm$  0.07 &  $-$  & 14.13 $\pm$  0.02 & 13.82 $\pm$  0.03 & 13.71 $\pm$  0.01 \\
 3 & 10:31:42.905 & -39:57:51.19 &  $-$  &  $-$  & 12.96 $\pm$  0.02 & 12.81 $\pm$  0.03 & 12.81 $\pm$  0.02 \\
 4 & 10:31:41.494 & -39:57:10.26 & 17.20 $\pm$  0.06 &  $-$  & 14.19 $\pm$  0.02 & 13.91 $\pm$  0.03 & 13.80 $\pm$  0.04 \\
 5 & 10:31:36.185 & -39:54:45.04 &  $-$  &  $-$  & 14.06 $\pm$  0.02 & 13.91 $\pm$  0.03 & 13.89 $\pm$  0.04 \\
 6 & 10:31:24.828 & -39:58:30.29 &  $-$  &  $-$  & 13.91 $\pm$  0.02 & 13.76 $\pm$  0.03 & 13.75 $\pm$  0.01 \\
 7 & 10:31:24.547 & -40:01:44.72 & 15.82 $\pm$  0.02 & 14.68 $\pm$  0.01 & 14.31 $\pm$  0.02 & 14.16 $\pm$  0.03 & 14.14 $\pm$  0.03 \\
 8 & 10:31:50.942 & -39:57:50.11 & 17.25 $\pm$  0.05 & 15.41 $\pm$  0.01 & 14.71 $\pm$  0.01 & 14.43 $\pm$  0.01 & 14.32 $\pm$  0.05 \\
 9 & 10:31:20.933 & -39:59:52.55 & 17.15 $\pm$  0.07 & 15.37 $\pm$  0.01 & 14.76 $\pm$  0.01 & 14.56 $\pm$  0.01 & 14.52 $\pm$  0.02 \\
 10 & 10:31:48.010 & -39:54:23.65 & 16.38 $\pm$  0.04 & 15.19 $\pm$  0.01 & 14.79 $\pm$  0.01 & 14.65 $\pm$  0.01 & 14.66 $\pm$  0.06 \\
 11 & 10:31:33.866 & -40:00:09.68 & 17.09 $\pm$  0.06 & 15.41 $\pm$  0.01 & 14.86 $\pm$  0.01 & 14.67 $\pm$  0.01 & 14.67 $\pm$  0.06 \\
 12 & 10:31:26.273 & -39:56:48.23 & 16.69 $\pm$  0.04 & 15.37 $\pm$  0.01 & 14.91 $\pm$  0.01 & 14.76 $\pm$  0.01 & 14.73 $\pm$  0.04 \\
 13 & 10:31:45.754 & -39:56:32.17 & 16.83 $\pm$  0.05 & 15.47 $\pm$  0.01 & 15.03 $\pm$  0.01 & 14.88 $\pm$  0.01 & 14.96 $\pm$  0.07 \\
 14 & 10:31:51.468 & -39:54:42.59 & 17.24 $\pm$  0.04 & 15.83 $\pm$  0.01 & 15.28 $\pm$  0.01 & 15.06 $\pm$  0.01 & 15.03 $\pm$  0.04 \\
 15 & 10:31:51.929 & -39:54:27.94 & 17.55 $\pm$  0.05 & 15.90 $\pm$  0.01 & 15.29 $\pm$  0.01 & 15.03 $\pm$  0.01 & 14.95 $\pm$  0.02 \\
 16 & 10:31:24.012 & -39:59:29.94 & 17.70 $\pm$  0.06 & 15.96 $\pm$  0.01 & 15.34 $\pm$  0.01 & 15.14 $\pm$  0.01 & 14.88 $\pm$  0.04 \\
 17 & 10:31:14.064 & -39:58:45.55 & 17.76 $\pm$  0.05 & 15.99 $\pm$  0.01 & 15.39 $\pm$  0.01 & 15.18 $\pm$  0.01 & 15.11 $\pm$  0.03 \\
 18 & 10:31:16.582 & -39:56:39.70 & 16.94 $\pm$  0.11 & 15.86 $\pm$  0.11 & 15.45 $\pm$  0.05 & 15.32 $\pm$  0.06 & 15.37 $\pm$  0.04 \\
 19 & 10:31:18.864 & -39:57:51.30 & 17.60 $\pm$  0.03 & 16.08 $\pm$  0.01 & 15.51 $\pm$  0.01 & 15.31 $\pm$  0.01 & 15.24 $\pm$  0.04 \\
 20 & 10:31:23.846 & -39:54:50.40 & 17.52 $\pm$  0.04 & 16.09 $\pm$  0.01 & 15.60 $\pm$  0.01 & 15.42 $\pm$  0.01 & 15.43 $\pm$  0.08 \\
 21 & 10:31:24.442 & -39:55:08.33 & 18.08 $\pm$  0.06 & 16.32 $\pm$  0.01 & 15.70 $\pm$  0.01 & 15.49 $\pm$  0.01 & 15.52 $\pm$  0.06 \\
 22 & 10:31:46.673 & -39:58:40.19 & 18.06 $\pm$  0.05 & 16.57 $\pm$  0.01 & 16.07 $\pm$  0.01 & 15.90 $\pm$  0.01 & 15.89 $\pm$  0.08 \\
 23 & 10:31:31.198 & -39:53:32.46 & 18.99 $\pm$  0.13 & 17.02 $\pm$  0.01 & 16.29 $\pm$  0.01 & 16.00 $\pm$  0.01 & 15.96 $\pm$  0.08 \\
 24 & 10:31:49.822 & -40:00:16.78 & 20.87 $\pm$  0.17 & 17.81 $\pm$  0.01 & 16.44 $\pm$  0.01 & 15.61 $\pm$  0.01 & 15.29 $\pm$  0.06 \\
 25 & 10:31:42.703 & -39:58:20.78 & 18.47 $\pm$  0.06 & 17.18 $\pm$  0.01 & 16.70 $\pm$  0.01 & 16.50 $\pm$  0.01 & 16.75 $\pm$  0.45 \\
 26 & 10:31:48.434 & -40:01:29.93 & 18.87 $\pm$  0.03 & 17.42 $\pm$  0.01 & 16.90 $\pm$  0.01 & 16.72 $\pm$  0.01 & 16.73 $\pm$  0.22 \\
 27 & 10:31:54.881 & -39:56:16.58 &  $-$  & 18.50 $\pm$  0.01 & 17.14 $\pm$  0.01 & 16.42 $\pm$  0.01 & 16.04 $\pm$  0.08 \\
 28 & 10:31:33.463 & -39:58:54.05 & 19.05 $\pm$  0.05 & 17.73 $\pm$  0.01 & 17.25 $\pm$  0.01 & 17.07 $\pm$  0.01 & 17.54 $\pm$  0.92 \\
 29 & 10:31:37.385 & -39:57:52.78 & 20.41 $\pm$  0.08 & 18.31 $\pm$  0.01 & 17.46 $\pm$  0.01 & 17.16 $\pm$  0.01 & 17.91 $\pm$  0.46 \\
 30 & 10:31:32.995 & -39:58:26.69 &  $-$  & 18.76 $\pm$  0.01 & 17.66 $\pm$  0.01 & 17.26 $\pm$  0.01 & 16.98 $\pm$  0.45 \\
 31 & 10:31:31.990 & -39:58:28.85 & 19.97 $\pm$  0.13 & 18.44 $\pm$  0.01 & 17.85 $\pm$  0.01 & 17.63 $\pm$  0.01 & 17.19 $\pm$  0.43 \\
\enddata
\end{deluxetable}



\begin{deluxetable}{cccccccc}
\tabletypesize{\scriptsize}
\tablecaption{$YJH$ photometry of the local photometric sequence}
\tablewidth{0pt}
\label{tab7.3}
\tablehead{
\colhead{id} & \colhead{R.A.} & \colhead{Dec.} & \colhead{$Y$} & \colhead{$J$} & \colhead{$H$}
}
\startdata
 1 & 10:31:45.355 & -40:00:42.73 & 12.23 $\pm$  0.01 & 11.99 $\pm$  0.01 & 11.70 $\pm$  0.01 \\
 2 & 10:31:44.107 & -39:55:25.68 & 12.78 $\pm$  0.01 & 12.57 $\pm$  0.01 & 12.34 $\pm$  0.02 \\
 3 & 10:31:42.905 & -39:57:51.19 & 13.03 $\pm$  0.01 & 12.69 $\pm$  0.01 & 12.19 $\pm$  0.01 \\
 4 & 10:31:41.494 & -39:57:10.26 & 13.11 $\pm$  0.01 & 12.76 $\pm$  0.01 & 12.29 $\pm$  0.02 \\
 5 & 10:31:36.185 & -39:54:45.04 & 13.21 $\pm$  0.01 & 12.97 $\pm$  0.01 & 12.68 $\pm$  0.01 \\
 6 & 10:31:24.828 & -39:58:30.29 & 13.22 $\pm$  0.01 & 12.85 $\pm$  0.01 & 12.37 $\pm$  0.02 \\
 7 & 10:31:24.547 & -40:01:44.72 & 13.55 $\pm$  0.01 & 13.27 $\pm$  0.01 & 12.92 $\pm$  0.01 \\
 8 & 10:31:50.942 & -39:57:50.11 & 13.41 $\pm$  0.01 & 13.16 $\pm$  0.01 & 12.87 $\pm$  0.03 \\
 9 & 10:31:20.933 & -39:59:52.55 & 13.69 $\pm$  0.01 & 13.32 $\pm$  0.01 & 12.75 $\pm$  0.01 \\
 10 & 10:31:48.010 & -39:54:23.65 & 13.59 $\pm$  0.01 & 13.29 $\pm$  0.01 & 12.86 $\pm$  0.01 \\
 11 & 10:31:33.866 & -40:00:09.68 & 13.73 $\pm$  0.01 & 13.45 $\pm$  0.01 & 13.14 $\pm$  0.01 \\
 12 & 10:31:26.273 & -39:56:48.23 & 13.71 $\pm$  0.01 & 13.35 $\pm$  0.01 & 12.84 $\pm$  0.01 \\
 13 & 10:31:51.468 & -39:54:42.59 & 13.83 $\pm$  0.01 & 13.42 $\pm$  0.01 & 12.88 $\pm$  0.01 \\
 14 & 10:31:51.929 & -39:54:27.94 & 13.97 $\pm$  0.01 & 13.69 $\pm$  0.01 & 13.39 $\pm$  0.02 \\
 15 & 10:31:24.012 & -39:59:29.94 & 14.05 $\pm$  0.01 & 13.75 $\pm$  0.01 & 13.42 $\pm$  0.01 \\
 16 & 10:31:14.064 & -39:58:45.55 & 14.15 $\pm$  0.01 & 13.90 $\pm$  0.01 & 13.59 $\pm$  0.01 \\
 17 & 10:31:16.582 & -39:56:39.70 & 14.28 $\pm$  0.01 & 14.00 $\pm$  0.01 & 13.71 $\pm$  0.01 \\
 18 & 10:31:18.864 & -39:57:51.30 & 14.21 $\pm$  0.01 & 13.80 $\pm$  0.01 & 13.28 $\pm$  0.01 \\
 19 & 10:31:23.846 & -39:54:50.40 & 14.30 $\pm$  0.01 & 14.05 $\pm$  0.01 & 13.76 $\pm$  0.01 \\
 20 & 10:31:24.442 & -39:55:08.33 & 14.24 $\pm$  0.01 & 14.00 $\pm$  0.01 & 13.73 $\pm$  0.02 \\
 21 & 10:31:46.673 & -39:58:40.19 & 14.29 $\pm$  0.01 & 13.97 $\pm$  0.01 & 13.58 $\pm$  0.02 \\
 22 & 10:31:31.198 & -39:53:32.46 & 14.39 $\pm$  0.01 & 14.10 $\pm$  0.01 & 13.71 $\pm$  0.02 \\
 23 & 10:31:49.822 & -40:00:16.78 & 14.48 $\pm$  0.01 & 14.02 $\pm$  0.01 & 13.35 $\pm$  0.01 \\
 24 & 10:31:42.703 & -39:58:20.78 & 14.54 $\pm$  0.01 & 14.24 $\pm$  0.01 & 13.90 $\pm$  0.01 \\
 25 & 10:31:48.434 & -40:01:29.93 & 14.77 $\pm$  0.01 & 14.49 $\pm$  0.01 & 14.15 $\pm$  0.01 \\
 26 & 10:31:54.881 & -39:56:16.58 & 14.74 $\pm$  0.01 & 14.23 $\pm$  0.01 & 13.67 $\pm$  0.01 \\
 27 & 10:31:33.463 & -39:58:54.05 & 15.09 $\pm$  0.01 & 14.82 $\pm$  0.01 & 14.48 $\pm$  0.02 \\
 28 & 10:31:37.385 & -39:57:52.78 & 15.25 $\pm$  0.01 & 14.95 $\pm$  0.01 & 14.71 $\pm$  0.01 \\
 30 & 10:31:32.995 & -39:58:26.69 & 15.72 $\pm$  0.01 & 15.24 $\pm$  0.01 & 14.70 $\pm$  0.01 \\
 29 & 10:31:31.990 & -39:58:28.85 & 15.63 $\pm$  0.01 & 15.12 $\pm$  0.01 & 14.57 $\pm$  0.02 \\
 31 & 10:31:21.082 & -39:58:30.44 & 15.84 $\pm$  0.01 & 15.56 $\pm$  0.01 & 15.30 $\pm$  0.03 \\
\enddata
\end{deluxetable}




\end{document}